\documentclass[10pt,letterpaper]{article}
\usepackage[top=0.85in,left=2.75in,footskip=0.75in]{geometry}

\usepackage{amsmath,amssymb}

\usepackage {tabularx}
\usepackage[utf8]{inputenc}
\usepackage{pgfplots}
\pgfplotsset{compat=1.18}
\usepackage{tikz}

\usepackage{graphicx}
\usepackage{subcaption}
\usepackage{array}
\usepackage[table]{xcolor}
\usepackage{float}

\usepackage{microtype}
\usepackage{ragged2e}
\usepackage{changepage}
\usepackage{verbatim}
\usepackage{textcomp,marvosym}

\usepackage{cite}

\usepackage{hyperref}
\usepackage{nameref}

\usepackage[right]{lineno}

\usepackage[
aboveskip=1pt,
labelfont=bf,
labelsep=period,
justification=raggedright,
singlelinecheck=off
]{caption}

\makeatletter
\renewcommand{\@biblabel}[1]{\quad#1.}
\makeatother

\usepackage{lastpage,fancyhdr}
\usepackage{epstopdf}
\fancyheadoffset[L]{2.25in}
\fancyfootoffset[L]{2.25in}
\raggedright
\title{Coordinated Information Dissemination on Telegram During Political Turbulence: A Case Study of Venezuela in Global News Channels}
\author{
Despoina Antonakaki$^{1,2}$,
Sotiris Ioannidis$^{2}$ \\
\\
$^{1}$Institute of Computer Science, Foundation for Research and Technology,\\ Vassilika Vouton, Heraklion, Crete, Greece \\
$^{2}$Technical University of Crete, University Campus, \\Akrotiri, Chania, Greece \\
}
\date{} % removes the date

\begin{document}

\maketitle

 \begin{abstract}
\justifying
Telegram has become a prominent venue for political communication and news dissemination, yet systematic evidence about coordinated content sharing on the platform remains limited. We investigate whether mainstream global news channels exhibit coordinated information dissemination when reporting on Venezuela during a prolonged period of political turbulence. Using a longitudinal collection of public Telegram posts from nine major international news outlets (2017--2026; 2{,}038 Venezuela-related messages), we operationalize coordination as the combination of (i) temporal synchronization (hourly and daily windows) and (ii) near-duplicate textual similarity measured via character $n$-gram TF--IDF cosine similarity.

Across both temporal resolutions, similarity scores are overwhelmingly concentrated at low values and we detect no cross-channel near-duplicate pairs at a conservative threshold of $\tau=0.85$, indicating that synchronized verbatim or near-verbatim reposting among mainstream outlets is rare in this setting. A negative control that randomly permutes timestamps within each channel yields the same null result across thresholds, confirming that the pipeline does not introduce spurious coordination signals.

We complement near-duplicate detection with three event-focused diagnostics. First, lead--lag analysis identifies temporal asymmetries in reporting, consistent with heterogeneous editorial responsiveness rather than synchronized copying. Second, narrative clustering during the peak window (3--6 January 2026) reveals moderate framing diversity (overall entropy $=1.87$), with channel-level asymmetries (e.g., lower entropy for France24 and broader coverage for RT) but no separable narrative blocs. Third, an Attention--Coordination Ratio (ACR) highlights extreme attention spikes in early January 2026 despite the absence of detected near-duplicate coordination, formalizing the distinction between attention synchronization and content coordination.

To contextualize Telegram dynamics with broader public attention, we also analyze an auxiliary Reddit subset; however, cross-community coordination is not estimable due to structural sparsity (no comparable daily buckets containing multiple subreddits). Overall, our results establish a conservative baseline: even during major geopolitical shocks that trigger simultaneous attention, mainstream news channels on Telegram tend to publish heterogeneous text rather than coordinated near-duplicate content.
\end{abstract}

\section{Introduction}
\justifying

Online messaging platforms have become central to the dissemination of political information, particularly during periods of political instability and international crisis. While platforms such as Twitter and Facebook have been extensively studied, Telegram remains comparatively underexplored, despite its rapid adoption by news organizations, political actors, and state-affiliated media. Telegram’s architecture---characterized by public channels, forwarding mechanisms, and limited content moderation---creates a distinct information environment that may facilitate synchronized dissemination of narratives at scale.

Venezuela represents a compelling case study for examining information dissemination during political turbulence. Over the past decade, the country has experienced sustained political instability, contested elections, international sanctions, and intense geopolitical attention. These developments have generated continuous global media coverage, often framed through competing political narratives. Understanding how such narratives propagate across platforms like Telegram is critical for assessing the dynamics of contemporary information ecosystems.

Existing research on political communication and information operations has largely focused on open social networks, particularly Twitter, where retweets, mentions, and follower networks provide rich signals for coordination analysis. However, Telegram differs substantially in both technical affordances and usage patterns. Content is primarily disseminated through channels rather than individual user interactions, and coordination may manifest through temporal synchronization and near-duplicate content rather than explicit network ties.

In this paper, we investigate whether and to what extent coordinated or synchronized information dissemination emerges in Telegram channels covering Venezuela-related news. Rather than assuming malicious intent, we adopt a conservative, data-driven approach that focuses on measurable coordination patterns, such as near-duplicate textual similarity and temporal alignment across channels.

We analyze a longitudinal dataset of public Telegram messages from major global news outlets, filtered to retain only content explicitly referring to Venezuela. Using a character $n$-gram--based similarity framework combined with temporal bucketing, we identify instances of cross-channel content alignment and examine their structural properties through network analysis.

Our findings indicate that near-duplicate coordination among mainstream global news channels is rare, even during major political events. When coordination does occur, it typically corresponds to breaking news events and reflects conventional news dissemination practices rather than sustained coordinated campaigns. These results establish an important baseline for understanding coordination on Telegram and highlight the necessity of distinguishing between normal journalistic synchronization and more strategic information operations.

\section{Contributions}
\justifying

This paper makes the following contributions:
\begin{itemize}
    \item We present one of the first empirical studies of political news dissemination on Telegram, focusing on global coverage of Venezuela during periods of sustained political turbulence.
    \item We propose a reproducible methodology for detecting coordinated information dissemination on Telegram based on character $n$-gram textual similarity and temporal synchronization, specifically designed for short news-style messages.
    \item We provide a longitudinal analysis of coordination patterns among major international news outlets on Telegram, establishing a baseline level of cross-channel content alignment in mainstream media ecosystems.
    \item We qualitatively validate detected coordination events and demonstrate that near-duplicate synchronization in mainstream outlets is rare and primarily associated with breaking news events rather than sustained coordinated campaigns.
    \item We release a transferable analytical framework that can be applied to more heterogeneous or partisan Telegram ecosystems to support future research on information operations and coordinated online behavior.
\end{itemize}

\section{Related Work}
\justifying

Research on political communication, coordination, and information dissemination in online environments has expanded significantly over the past decade. While early work has predominantly focused on open social media platforms such as Twitter, recent studies increasingly recognize the importance of messaging platforms---particularly Telegram---as influential spaces for political discourse, mobilization, and news dissemination.

Twitter has long served as a primary platform for computational social science research due to its accessible data model and interaction structure. A comprehensive survey by Antonakaki et al.~\cite{antonakaki2021survey} systematically maps Twitter research across social graph analysis, sentiment and content analysis, and the detection of malicious activities such as bots and coordinated attacks. Several studies examine political turbulence through large-scale Twitter analysis \cite{antonakaki2017turbulence}, while related work on Venezuela highlights polarized discussions and mobilization dynamics during crisis periods \cite{drivers_polarization_venezuela,social_media_mobilization_venezuela}. Coordination and manipulation on Twitter have been studied using approaches based on trending-topic abuse \cite{antonakaki2016trending}, network growth properties \cite{antonakaki2018degree}, and explainable machine learning for bot detection \cite{shevtsov2022explainable,shevtsov2022discovery,antonakaki2023botartist}.

Compared to Twitter, Telegram has received more limited but growing scholarly attention. Prior work frames Telegram as a political communication channel used by elites and political actors \cite{salikov2019telegram}, while methodological studies emphasize the platform’s research opportunities and the feasibility of content and network analysis using Telegram data \cite{khaund2021telegramdata}. Telegram’s role in political campaigns has been explored through content analysis of party communication strategies \cite{alonso2022telegramcampaign}. Work on coordination and mobilization on Telegram includes large-scale studies of election-related coordination, where dissemination networks and key relays are identified using network extraction and text analysis \cite{venancio2024telegramcoord}. Telegram has also been examined as a news dissemination space and as a form of alternative media infrastructure \cite{alrawi2022news}, while studies of journalistic discourse highlight how national and international media construct narratives on social platforms during Venezuela-related crises \cite{afrouzi2020venezuela,dronova2021venezuela}.

Building on these research directions, our study focuses on coordination signals in Telegram news channels via conservative measures of temporal synchronization and near-duplicate textual similarity, aiming to establish a baseline for mainstream global news coverage about Venezuela.

\section{Data Collection and Preprocessing}
\justifying

To the best of our knowledge, no publicly available, curated dataset exists that captures Venezuela-related political news dissemination and coordination across social media platforms. Existing datasets are either platform-specific, temporally limited, focused on user-level interactions, or oriented toward sentiment and engagement rather than synchronized content reuse. Moreover, access restrictions and data deletions on major platforms such as Twitter/X significantly limit the reproducibility of longitudinal studies.

Telegram therefore constitutes a uniquely suitable platform for this analysis. Its public broadcast channels, stable archival access, and message-level temporal transparency enable systematic, long-term observation of news dissemination behavior without reliance on engagement signals or proprietary recommendation mechanisms.

We collected public Telegram messages from official channels operated by major international news organizations using the Telegram API and a custom-built incremental crawler. The crawler retrieves messages chronologically, applies conservative rate limiting, and stores records in append-only JSONL files while maintaining per-channel state files. This design prevents duplication and ensures fault-tolerant data acquisition across repeated collection runs.

The dataset includes Telegram channels from BBC World, CNN, Reuters, RT, Al Jazeera, France~24, DW, Euronews, Sky News, and the Associated Press. These channels were selected to represent mainstream global news coverage with sustained international reach. Messages were filtered to retain only content explicitly referencing Venezuela, using a curated set of country-, location-, and actor-specific keywords. This conservative filtering strategy prioritizes precision over recall, ensuring that retained messages are directly relevant to the case study.

\begin{figure}[h]
    \centering
    \includegraphics[width=0.6\linewidth]{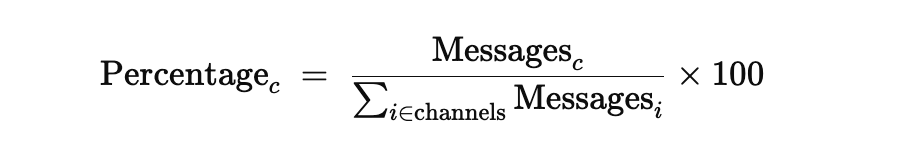}
    \caption{Relative contribution of Venezuela-related messages per Telegram channel.}
    \label{fig:channel_percentage}
\end{figure}

\begin{table}[h]
\centering
\caption{Number and percentage of Venezuela-related messages per Telegram channel.}
\label{tab:messages_per_channel}
\begin{tabular}{lrr}
\hline
\textbf{Channel} & \textbf{Messages} & \textbf{Percentage (\%)} \\
\hline
RT News & 1461 & 71.37 \\
BBC World & 310 & 15.14 \\
France 24 & 178 & 8.70 \\
BBC World (feed) & 73 & 3.57 \\
Euronews & 12 & 0.59 \\
Associated Press & 8 & 0.39 \\
DW News & 3 & 0.15 \\
Sky News & 1 & 0.05 \\
CNN Breaking & 1 & 0.05 \\
\hline
\end{tabular}
\end{table}

Text preprocessing was intentionally minimal. Messages were lowercased and lightly normalized by removing URLs and excessive whitespace, while preserving punctuation and stylistic features. We deliberately avoided aggressive linguistic preprocessing such as stemming or lemmatization in order to retain lexical and formatting cues that are informative for near-duplicate detection in short news-style texts.

\begin{table}[h]
\centering
\caption{Summary statistics of the Venezuela-related Telegram dataset.}
\label{tab:dataset_summary}
\begin{tabular}{l r}
\hline
\textbf{Statistic} & \textbf{Value} \\
\hline
Number of channels & 9 \\
Total messages (Venezuela-related) & 2038 \\
Date range & 2017--2026 \\
Median messages per channel & 12 \\
Mean messages per channel & 227.4 \\
\hline
\end{tabular}
\end{table}

\subsection{Auxiliary Reddit dataset}
\justifying
To complement Telegram-based findings with an auxiliary public-discussion context, we additionally collect Reddit submissions and comment threads from multiple Venezuela-focused subreddits. This Reddit dataset is not used for the core cross-channel Telegram coordination detection results; rather, it is intended to support contextual analyses (e.g., identifying concurrent public attention peaks, validating event timelines, and enabling future cross-platform comparisons).
 
We target the following subreddits:
\begin{itemize}
    \item \texttt{r/venezuela}
    \item \texttt{r/VenezuelaPolitics}
    \item \texttt{r/AskVenezuela}
    \item \texttt{r/venezuelancivilwar}
    \item \texttt{r/venezuela\_is\_not\_iraq}
    \item \texttt{r/VenezuelaUpdate} (empty during the collection period)
\end{itemize}

 \begin{table}[H]
\centering
\caption{Summary statistics of Reddit datasets used in the analysis.}
\label{tab:reddit_dataset_summary}
\footnotesize
\begin{tabularx}{\linewidth}{p{3.4cm}rrrr}
\hline
\textbf{Dataset} &
\textbf{Posts} &
\textbf{Rep. Comm.} &
\textbf{Downl. Comm.} &
\textbf{Coverage (\%)} \\
\hline
\texttt{r/venezuela} &
929 &
18{,}258 &
0 &
0.0 \\
\texttt{r/VenezuelaPolitics} &
0 &
0 &
0 &
-- \\
\texttt{r/worldnews} (Venezuela-related) &
182 &
98{,}866 &
36{,}244 &
36.7 \\
\hline
\textbf{Total} &
1{,}111 &
117{,}124 &
36{,}244 &
30.9 \\
\hline
\end{tabularx}
\end{table}

In addition, we collect from a generic, high-volume news subreddit:
\begin{itemize}
    \item \texttt{r/worldnews} (restricted to submissions that mention Venezuela, using a keyword-based filter applied to titles, selftext, and URLs)
\end{itemize}

Data are collected from publicly accessible Reddit endpoints in a manner consistent with platform rate limits and basic ethical constraints. For each submission, we store lightweight metadata (e.g., subreddit, title, author, timestamp, URL, score, and number of comments). For each submission, we also retrieve the associated comment tree and store individual comments with their timestamps and parent-child relationships, enabling thread-level reconstruction when feasible.
\paragraph{Reddit comment coverage.}
For subreddit-specific datasets "venezuela",  and "VenezuelaPolitics,  only submission-level metadata were collected, and no comments were downloaded. In contrast, for the generic \texttt{r/worldnews} subreddit, we collected full discussion threads for posts mentioning Venezuela, including all comments returned by Reddit’s public JSON endpoints. While Reddit reports a total of 98{,}866 comments across these posts, we successfully downloaded 36{,}244 comment records (36.7\%). The discrepancy is expected and arises from limitations of unauthenticated Reddit endpoints, including truncated responses for large threads, deleted or removed comments, and the presence of \emph{``more''} placeholders that require additional API calls. As a result, the collected comments represent a best-effort subset of the full discussion while preserving large-scale conversational structure.

\subsection{Preprocessing}
\justifying

 \justifying
Text preprocessing was intentionally minimal across datasets in order to preserve stylistic, lexical, and formatting cues that are informative for near-duplicate detection and temporal alignment analyses.

\paragraph{Telegram normalization.}
Telegram messages were lowercased and lightly normalized by (i) removing URLs, (ii) collapsing repeated whitespace, and (iii) stripping non-informative control characters, while preserving punctuation, emojis, and other surface markers. We deliberately avoided aggressive linguistic preprocessing (e.g., stemming, lemmatization, or stopword removal) because such transformations can reduce discriminative signals in short, headline-style posts and can obscure near-duplicate reuse patterns.

\paragraph{Reddit normalization and thread structure.}
Reddit submissions and comments were stored largely in raw form (with minimal normalization for encoding consistency and whitespace) to preserve conversational structure and pragmatic markers. For thread-based collections, each record includes identifiers that enable reconstruction of comment trees (submission ID, comment ID, parent ID, and depth when available). This design supports downstream thread-level analyses (e.g., temporal discussion bursts, reply cascades, and qualitative inspection) without altering the original text.

\paragraph{Storage format.}
All datasets are stored in line-delimited JSON (JSONL). Each record contains lightweight metadata (timestamp, author/channel identifier, source, and URL/permalink when available), enabling reproducible filtering and aggregation without dependence on platform-specific engagement signals.

\subsection {Platform-Specific Feasibility of Coordination Analysis}

We attempted to replicate the Telegram coordination detection framework on Reddit by computing cross-subreddit near-duplicate similarity within fixed temporal windows. Using 246 Venezuela-related submissions spanning 67 daily buckets, we find that no bucket contains submissions from more than one subreddit, yielding zero comparable cross-source pairs. Consequently, cross-community near-duplicate synchronization under shared temporal windows is not estimable on Reddit due to structural sparsity and asynchronous posting behavior. We therefore use Reddit exclusively to contextualize attention dynamics (volume and temporal bursts) rather than as a substrate for coordination detection.
\begin{figure}[h]
\centering
\begin{tikzpicture}
\begin{axis}[
    ybar,
    bar width=25pt,
    ymin=0,
    ylabel={Comparable temporal buckets},
    symbolic x coords={Telegram, Reddit},
    xtick=data,
    nodes near coords,
    nodes near coords align={vertical},
    width=0.55\linewidth,
    height=5cm,
    enlarge x limits=0.4
]
\addplot coordinates {(Telegram, 120) (Reddit, 0)};
\end{axis}
\end{tikzpicture}
\caption{Number of temporal buckets containing content from at least two distinct sources, enabling cross-source coordination analysis. Telegram exhibits numerous comparable buckets, whereas Reddit exhibits none due to structural sparsity and asynchronous posting.}
\label{fig:coord_feasibility}
\end{figure}
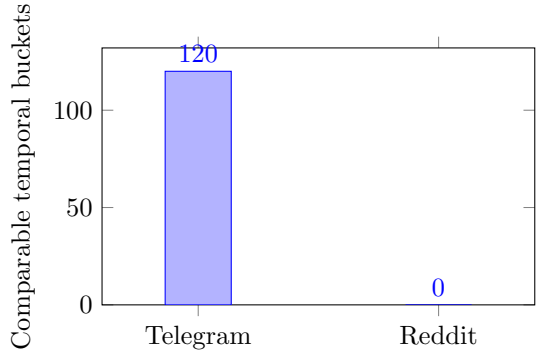

\begin{comment}
\begin{table}[h]
\centering
\caption{Feasibility of cross-source coordination analysis across platforms.}
\label{tab:coord_feasibility}
\begin{tabular}{lcccc}
\hline
\textbf{Platform} & \textbf{Items} & \textbf{Buckets} & \textbf{Comparable buckets} & \textbf{Coordination feasible} \\
\hline
Telegram & 2{,}038 & Daily / Hourly & Yes & \checkmark \\
Reddit & 246 & Daily & 0 & $\times$ \\
\hline
\end{tabular}
\end{table}

\end{comment}

\begin{table}[ht]
    \centering
    \caption{Feasibility of cross-source coordination analysis across platforms.}
    \label{tab:coord_feasibility}
    \small
    \setlength{\tabcolsep}{4pt}
    \renewcommand{\arraystretch}{1.1}
    
    \resizebox{\linewidth}{!}{%
    \begin{tabular}{@{}l r l c c@{}}
    \hline
    \textbf{Platform} & \textbf{Items} & \textbf{Buckets} & \textbf{Comp.buckets} & \textbf{Coord.feasible} \\
    \hline
    Telegram & 2{,}038 & Daily / Hourly & Yes & Yes \\
    Reddit   & 246     & Daily          & 0   & No  \\
    \hline
    \end{tabular}%
    }
    \end{table}

Figure~\ref{fig:coord_feasibility} illustrates the structural preconditions for coordination detection across platforms. While Telegram routinely exhibits overlapping temporal activity across channels, Reddit submissions remain isolated within single subreddits, preventing cross-community synchronization analysis.

\paragraph{Negative control on Reddit (timestamp randomization).}
To mirror the Telegram falsification test, we additionally randomized timestamps within each subreddit (analogous to within-channel shuffling on Telegram) and re-ran the cross-source coordination pipeline. The result is unchanged: both the original and randomized data contain 67 daily buckets and zero comparable buckets (i.e., no day contains submissions from more than one subreddit), yielding zero cross-subreddit pairs above threshold for all $\tau$. This demonstrates that, in this Reddit setting, timestamp randomization is structurally uninformative because the limiting factor is cross-community sparsity rather than temporal alignment.
In our data, the pipeline yields \emph{total\_buckets}=67, \emph{comparable\_buckets}=0, and \emph{total\_pairs}=0 both before and after timestamp randomization.
\begin{table}[h]
\centering
\caption{Reddit negative control (timestamp randomization within subreddit).}
\label{tab:reddit_neg_control}
\begin{tabular}{lrrrr}
\hline
\textbf{Condition} & \textbf{Buckets} & \textbf{Comp. buck.} & \textbf{Pair buck.} & \textbf{Pairs ($\tau=0.85$)} \\
\hline
Original & 67 & 0 & 0 & 0 \\
Shuffl.timst. & 67 & 0 & 0 & 0 \\
\hline
\end{tabular}
\end{table}

In addition, sensitivity analysis over similarity thresholds and timestamp randomization experiments analogous to those performed on Telegram yield trivial outcomes on Reddit: for all thresholds $\tau$, the number of cross-subreddit pairs remains zero, and randomizing timestamps within subreddits leaves results unchanged. This reflects a collapse of the coordination parameter space due to structural sparsity rather than evidence of absent coordination.
Detailed auxiliary Reddit experiments corresponding to sensitivity analysis, network projection, and attention-versus-coordination analysis are reported in Appendix~A.

\section{Volume and Temporal Characteristics}
\justifying

Before examining coordinated information dissemination, we analyze the volume and temporal distribution of Venezuela-related messages across Telegram news channels. This descriptive analysis provides essential context for understanding baseline publication behavior and the conditions under which synchronization may occur.

Message volumes vary substantially across channels, reflecting differences in editorial focus and publishing frequency. A small subset of channels accounts for a disproportionate share of Venezuela-related content, while others contribute sporadically. Temporal analysis reveals a bursty publishing pattern: long periods of low baseline activity are interspersed with sharp peaks corresponding to major political events such as elections, protests, or international policy announcements. Inter-arrival time distributions indicate that most messages are separated by relatively long intervals, while a small fraction are published within short time windows during breaking news cycles.

These volume and temporal characteristics have direct implications for coordination analysis. High-volume channels are statistically more likely to produce near-synchronous messages simply due to publishing frequency. Consequently, coordination signals must be interpreted relative to baseline activity levels rather than isolated coincidences.

\begin{figure}[h]
    \centering
    \includegraphics[width=0.55\linewidth]{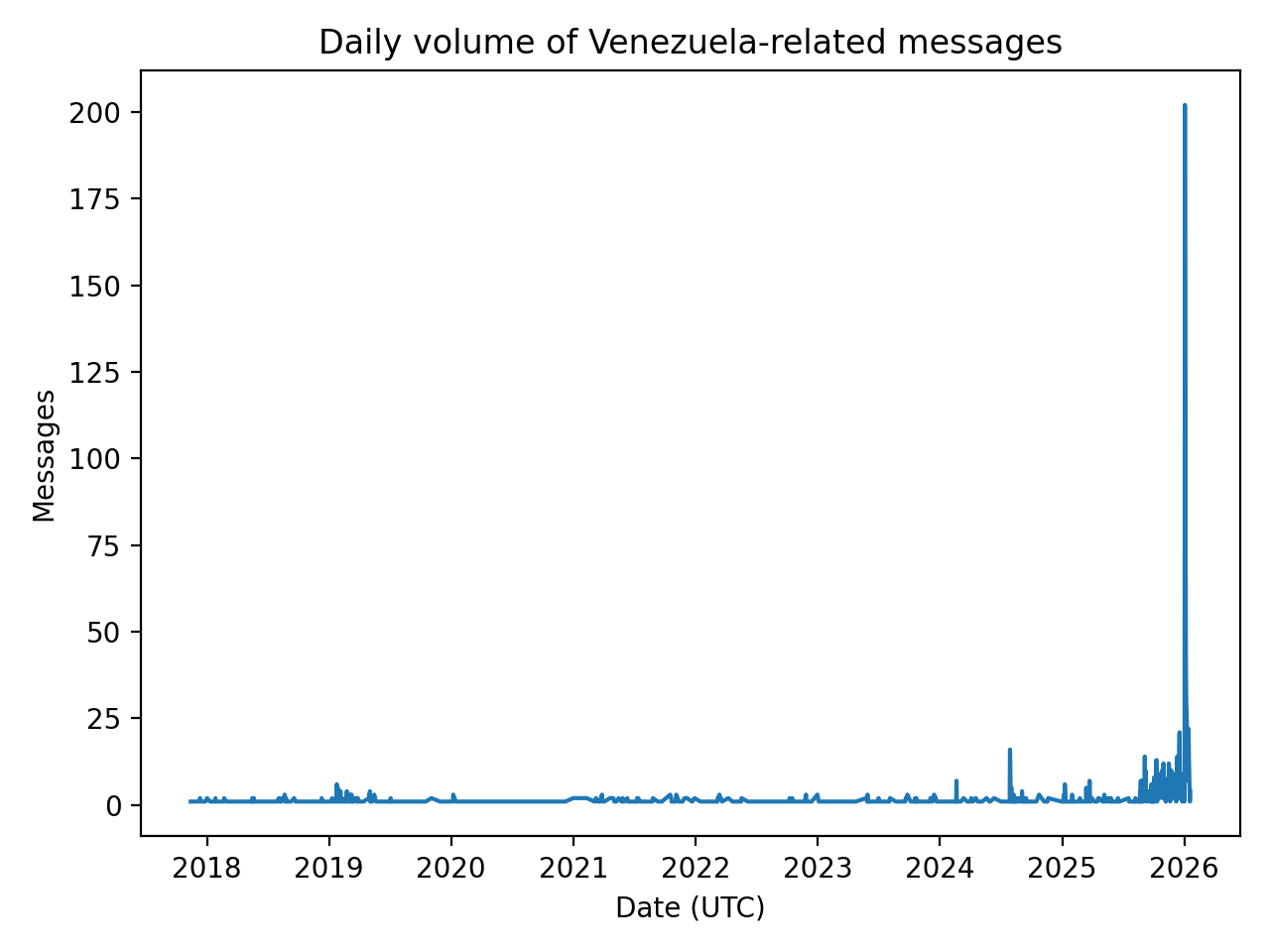}
    \caption{Daily volume of Venezuela-related messages across the monitored Telegram channels.}
    \label{fig:fig_daily_volume}
\end{figure}

\begin{figure}[h]
    \centering
    \includegraphics[width=0.55\linewidth]{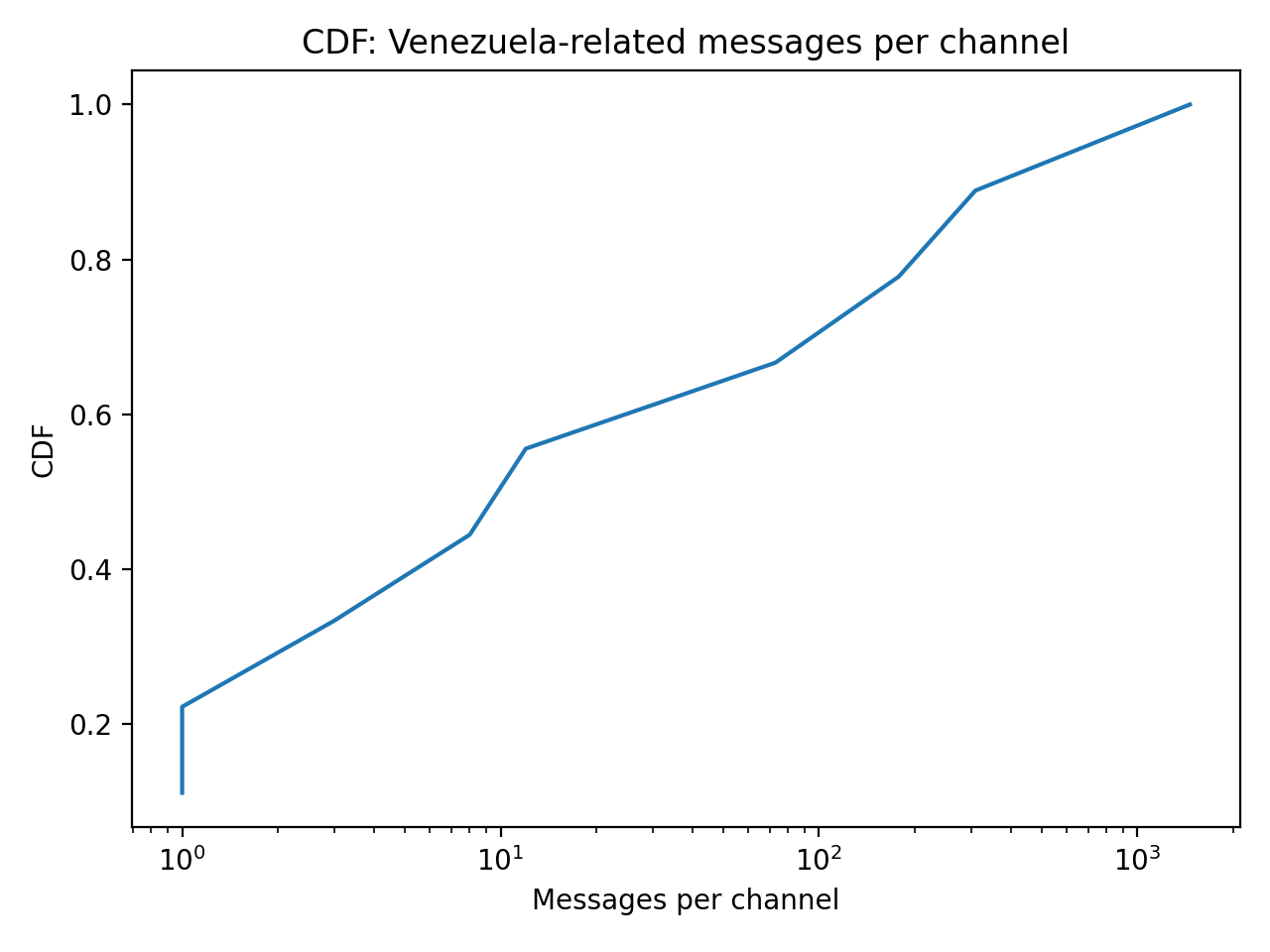}
    \caption{CDF of Venezuela-related message counts per channel, highlighting a highly skewed distribution.}
    \label{fig:fig_cdf_messages_per_channel}
\end{figure}

\subsection{Reddit volume analysis}
\label{subsec:reddit_volume}

To contextualize Telegram activity with broader public discussion dynamics, we analyze daily Reddit submission volume related to Venezuela across both subreddit-specific communities and a global news forum. Importantly, for comparability across subreddits with heterogeneous data availability, we restrict this analysis to \emph{submission-level activity only}, excluding comments. This choice avoids distortions arising from uneven comment coverage across datasets and ensures that all series reflect the same unit of activity.
 \begin{figure}
     \centering
     \includegraphics[width=0.5\linewidth]{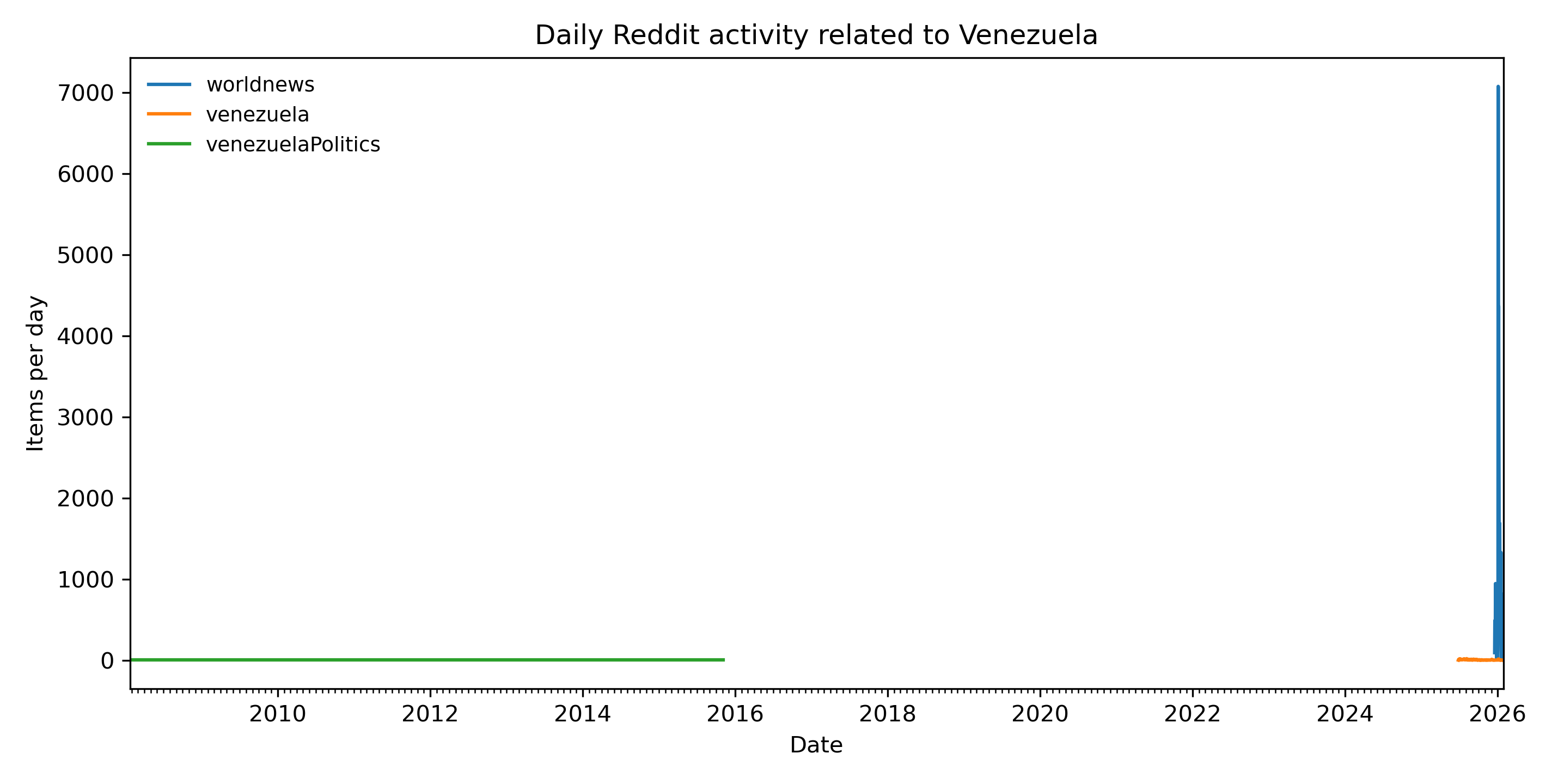}
     \caption{Daily Reddit activity related to Venezuela across multiple subreddits. 
The figure shows the number of items (submissions and comments) per day for Venezuela-focused subreddits and for \texttt{r/worldnews} restricted to Venezuela-related posts. 
While Venezuela-specific communities exhibit relatively low and stable activity, \texttt{r/worldnews} displays pronounced, short-lived spikes corresponding to major geopolitical events that generate large discussion threads. 
These peaks reflect event-driven public attention rather than coordinated content dissemination.}

     \label{fig:placeholder}
 \end{figure}
\begin{figure}[h]
    \centering
    \includegraphics[width=0.6\linewidth]{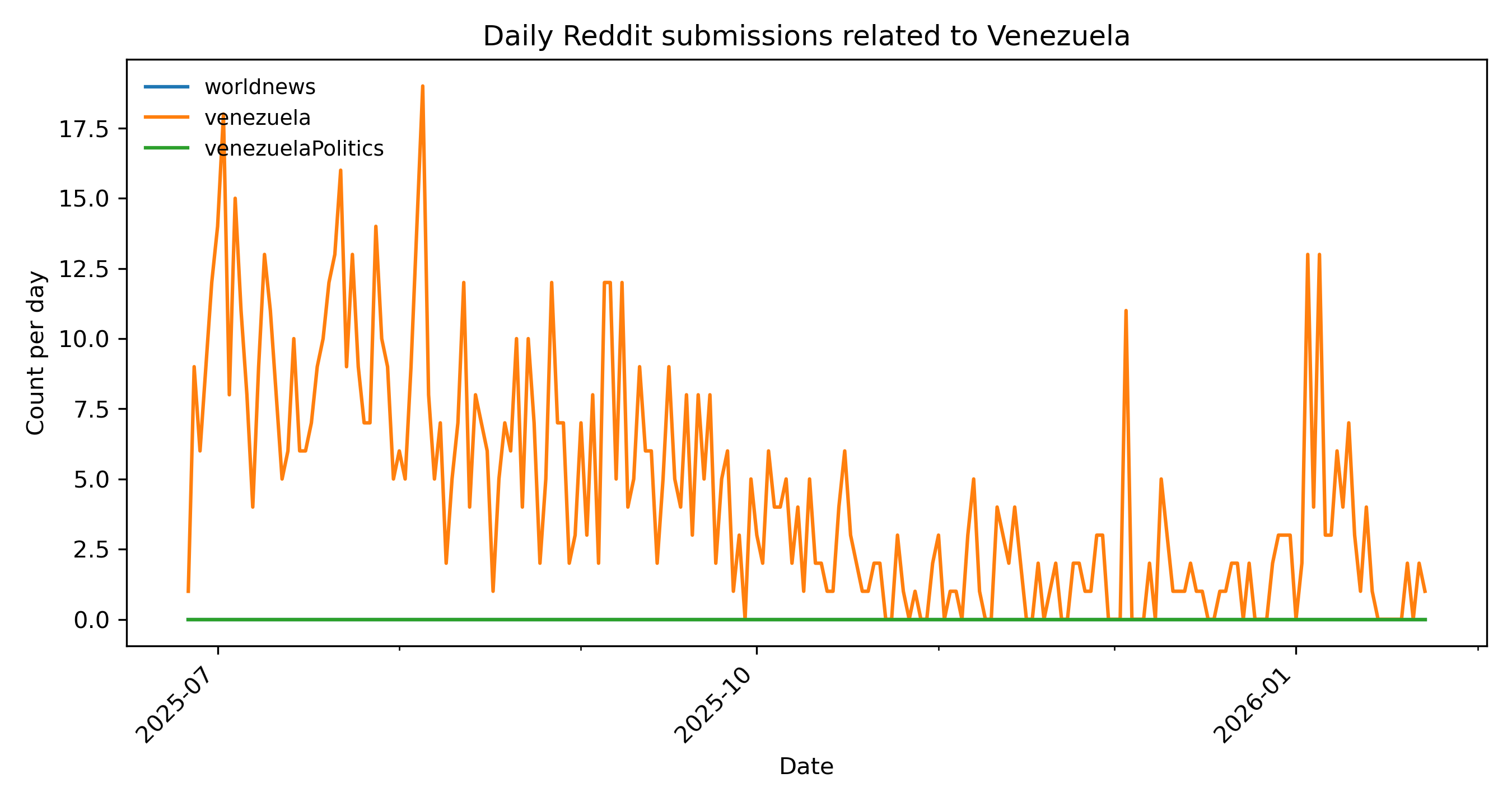}
    \caption{Daily Reddit submission volume related to Venezuela across subreddit-specific and global news communities. Only submission-level activity is shown. The dominance of \texttt{r/venezuela} reflects localized agenda-setting, while other subreddits exhibit negligible submission activity during this period.}
    \label{fig:reddit_volume_posts}
\end{figure}

Fig.~\ref{fig:reddit_volume_posts} shows the daily number of Venezuela-related Reddit submissions for \texttt{r/venezuela},
and \texttt{r/VenezuelaPolitics}, and Venezuela-related posts in \texttt{r/worldnews}. The figure reveals that almost all submission activity during the observation window originates from \texttt{r/venezuela}. In contrast, \texttt{r/worldnews} and \texttt{r/VenezuelaPolitics} exhibit zero or negligible submission activity in this period, resulting in flat baselines at zero.

This asymmetric pattern reflects structural differences in subreddit usage rather than data artifacts. Venezuela-focused subreddits primarily serve as venues for original submissions and discussion initiation, whereas global news communities such as \texttt{r/worldnews} predominantly host high-engagement comment threads attached to a relatively small number of submissions. Because comment activity is intentionally excluded here, the resulting volume plot highlights where new discussion threads are initiated rather than where discussion intensity is highest.

The observed dominance of \texttt{r/venezuela} in submission volume indicates that Reddit-based agenda-setting related to Venezuela during this period is concentrated within country-specific communities rather than global news forums. This finding complements the Telegram analysis by showing that peaks in attention on Reddit, like those on Telegram, are driven by localized editorial or community dynamics rather than synchronized cross-community posting behavior.

\begin{figure}
    \centering
    \includegraphics[width=0.9\linewidth]{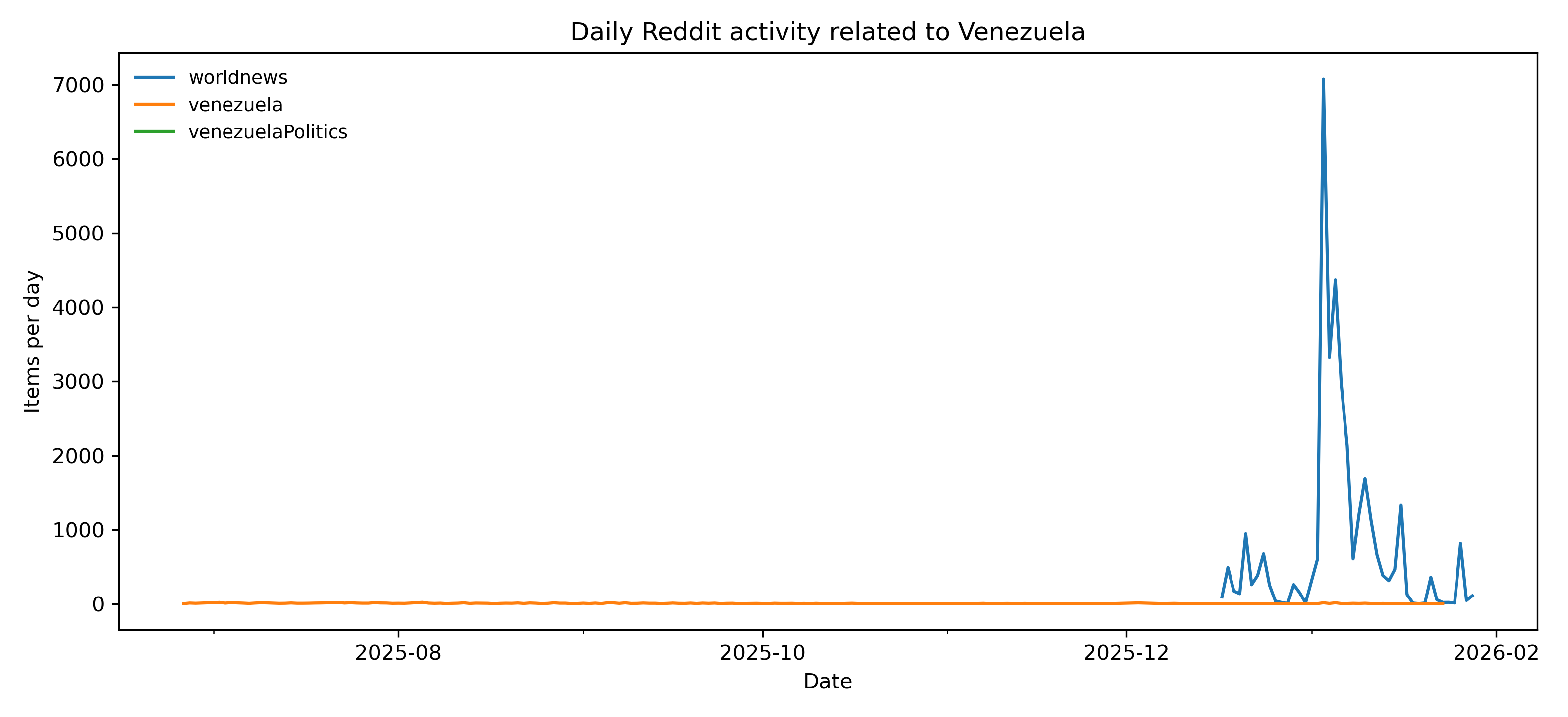}
    \caption{Daily Reddit activity related to Venezuela across multiple subreddits.}
    \label{fig:reddit_acrosssubre}
\end{figure}

The Reddit time series are constructed using each record’s publication timestamp (created\_utc) rather than download time, and the analysis is zoomed to the 2025–2026 period to highlight event-driven discussion bursts around the January 2026 escalation \ref{fig:reddit_acrosssubre}.

\subsection{Reddit cross-source similarity (auxiliary)}
\label{subsec:reddit_similarity}

\begin{figure}
    \centering
    \includegraphics[width=0.5\linewidth]{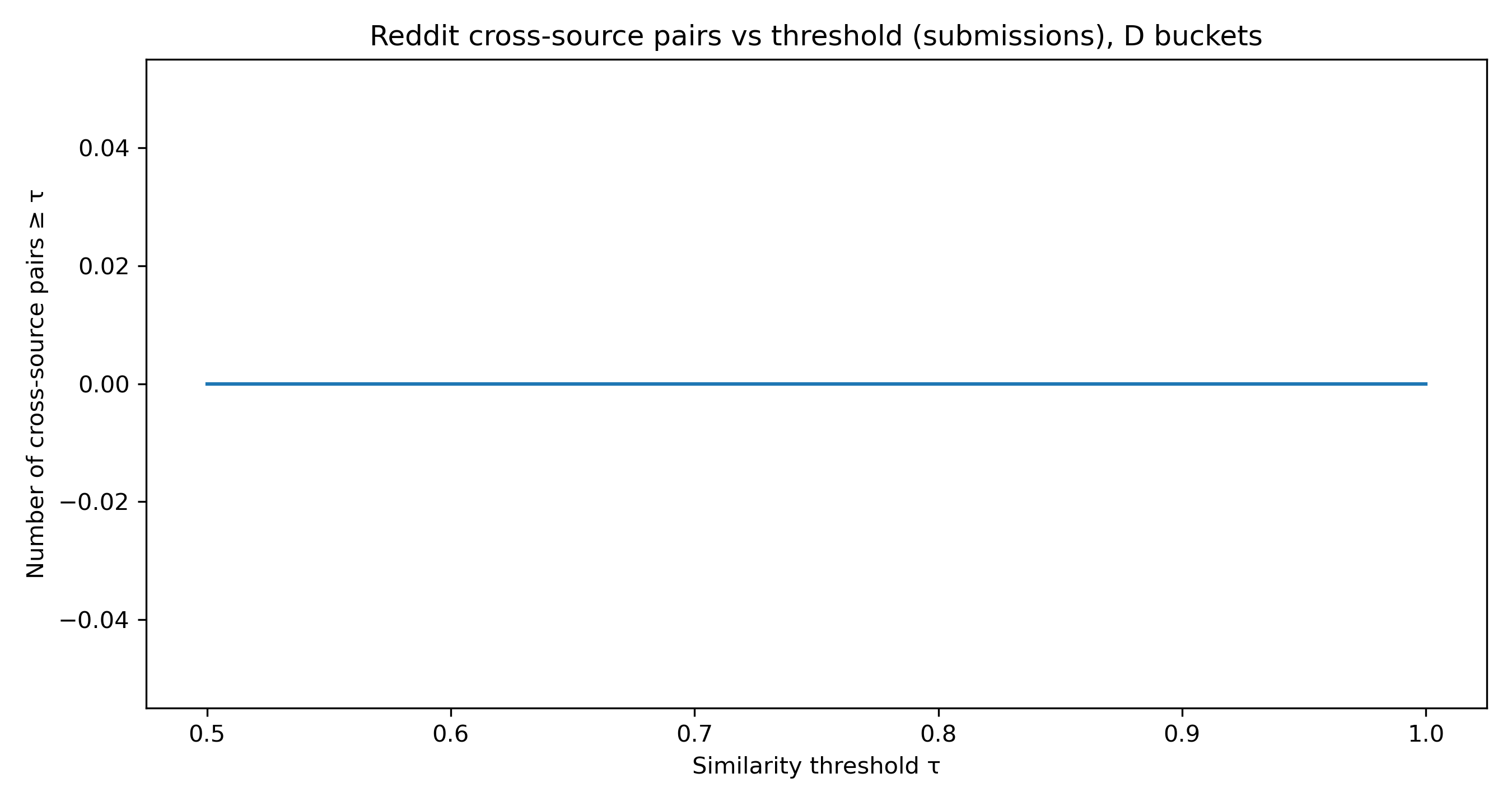}
\caption{Sensitivity curve on Reddit (daily buckets). The number of detected cross-subreddit near-duplicate pairs remains zero for all thresholds due to the absence of comparable buckets.}
\label{fig:reddit_pairs_D}

\end{figure}

\begin{figure}
    \centering
    \includegraphics[width=0.5\linewidth]{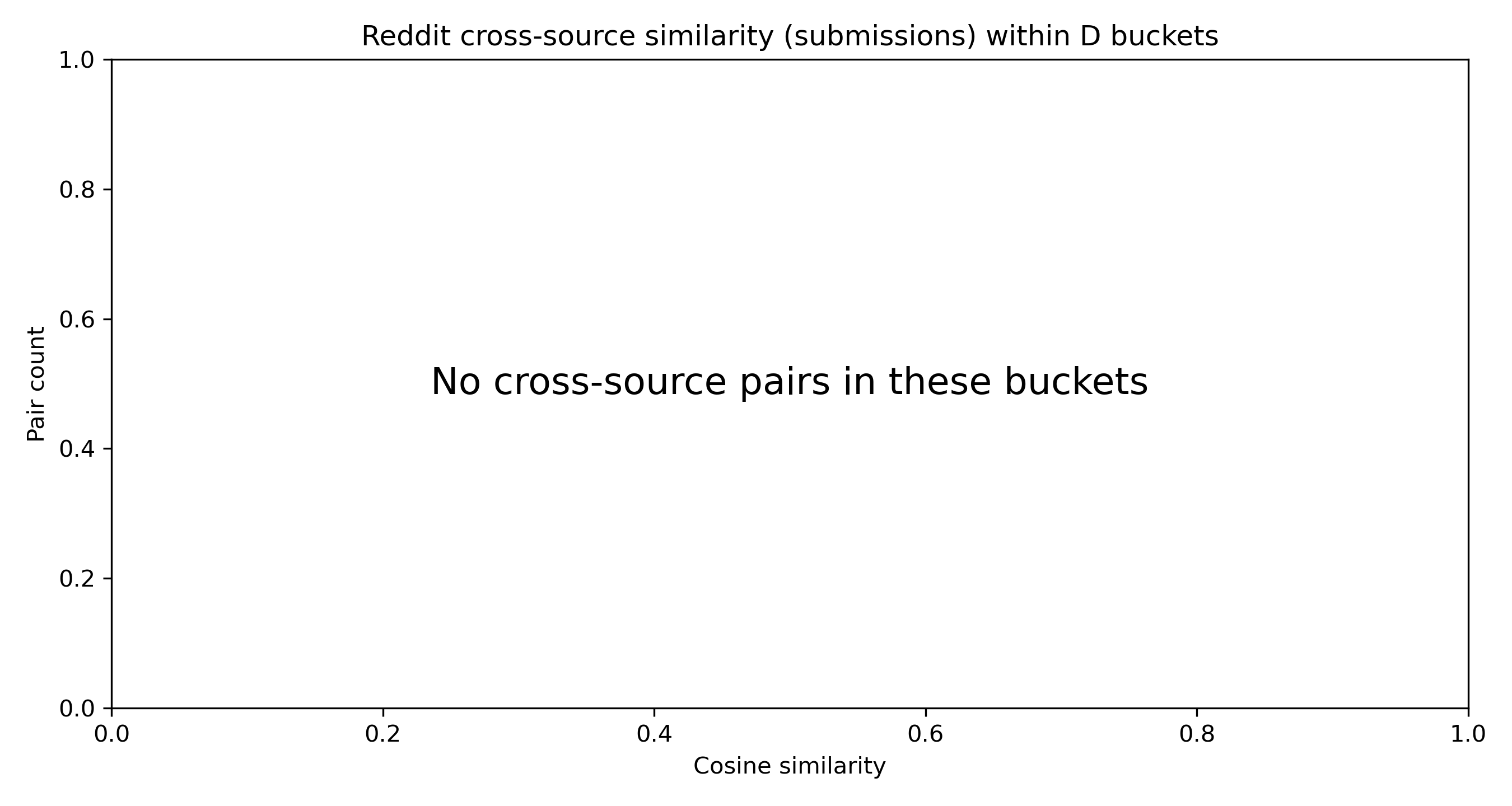}
\caption{Reddit similarity distribution is not estimable under daily bucketing because no cross-subreddit submission pairs exist within the same temporal windows.}
\label{fig:reddit_sim_hist_D}

\end{figure}

We attempted an auxiliary cross-source similarity analysis across Reddit communities by computing character $n$-gram TF--IDF cosine similarity between \emph{submissions} originating from different subreddits within fixed time buckets. In contrast to the Telegram setting, this analysis yields no cross-source submission pairs within the same daily buckets, and therefore no similarity distribution can be estimated (Fig.~\ref{fig:reddit_sim_hist_D}) and the number of pairs above threshold remains zero for all $\tau$ (Fig.~\ref{fig:reddit_pairs_D}).

This null result is driven by sparsity and asynchronous posting across the selected subreddits: during the observation window, most days contain submissions from only one source (primarily \texttt{r/venezuela}), while other sources contribute zero or negligible submission activity. Consequently, cross-community near-duplicate detection on submissions is not informative in this auxiliary Reddit setting. We therefore use Reddit primarily as contextual evidence of attention dynamics (volume) rather than as a coordination substrate analogous to Telegram channels.


 \section{Coordination Detection Methodology}
\justifying

To identify coordinated information dissemination on Telegram, we adopt a conservative, content-based approach that combines temporal synchronization and near-duplicate textual similarity. This design choice reflects both the broadcast-oriented architecture of Telegram channels and the journalistic nature of mainstream news content, where alignment may emerge through near-simultaneous reporting of similar information rather than explicit interaction links.

Messages are grouped into fixed temporal windows (hourly or daily) based on publication timestamps. Within each window, we compute pairwise textual similarity only for messages posted by different channels. Textual similarity is measured using a character $n$-gram TF--IDF representation and cosine similarity, which is robust to minor differences in punctuation, formatting, and wording typical of short news-style messages. Pairs exceeding a chosen similarity threshold $\tau$ are flagged as near-duplicate, temporally aligned instances and are treated as coordination candidates.

\section{Coordination Analysis Results}
\label{sec:coord_results}
\justifying

This section reports the empirical results of the coordination analysis under different temporal resolutions. We focus on the distribution of textual similarity between cross-channel message pairs and the sensitivity of detected coordination events to the similarity threshold.

\subsection{Hourly Coordination ($\texttt{h}$ buckets)}

\begin{figure}[H]
    \centering
    \includegraphics[width=0.7\linewidth]{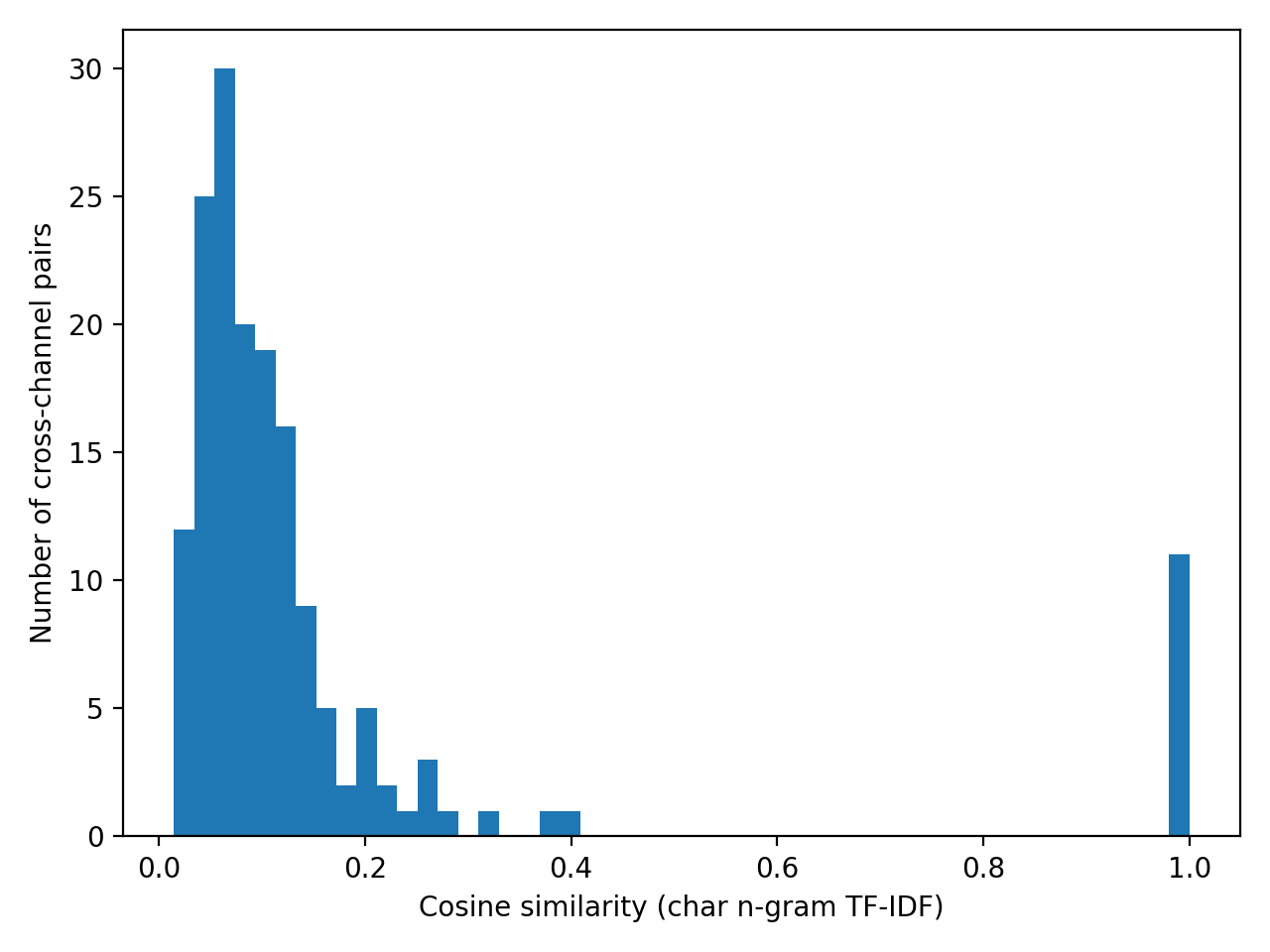}
    \caption{Distribution of cross-channel cosine similarities within hourly time buckets ($\texttt{h}$). Most pairs exhibit low similarity, with only a negligible number of near-duplicate cases.}
    \label{fig:sim_hist_h}
\end{figure}

We first analyze coordination using hourly time buckets, which capture near-simultaneous publication behavior. Fig.~\ref{fig:sim_hist_h} shows the distribution of cosine similarity scores between cross-channel message pairs within the same hourly bucket.

The distribution is heavily concentrated at low similarity values, indicating that most messages published within the same hour are not textually aligned across channels. Only a very small number of pairs exhibit extremely high similarity (near 1.0), corresponding to rare cases of near-verbatim reuse.

\begin{figure}[H]
    \centering
    \includegraphics[width=0.7\linewidth]{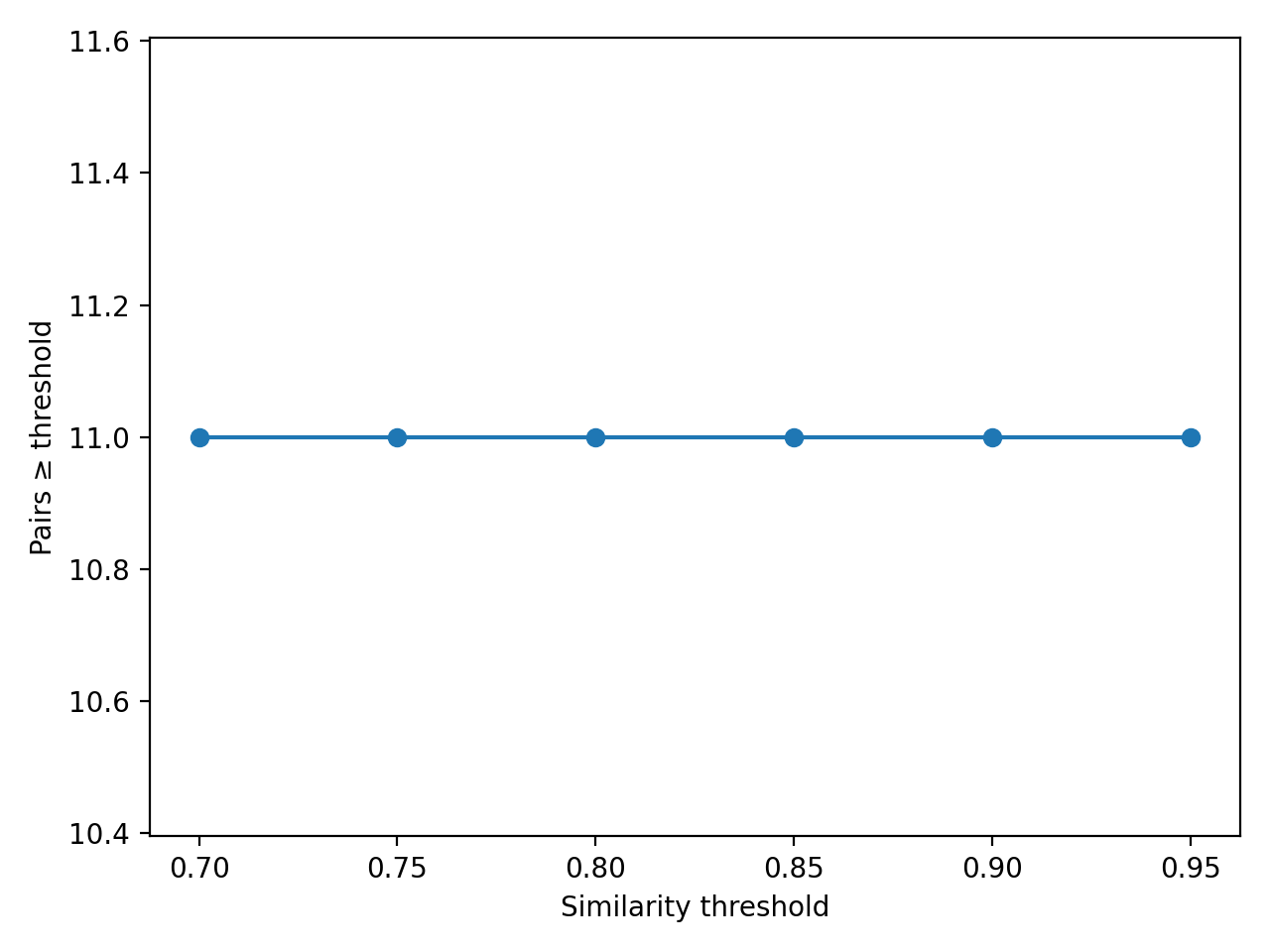}
    \caption{Number of detected cross-channel pairs as a function of the similarity threshold under hourly bucketing. No pairs exceed strict thresholds, indicating the absence of tight synchronization.}
    \label{fig:pairs_vs_threshold_h}
\end{figure}

Fig.~\ref{fig:pairs_vs_threshold_h} reports the number of detected cross-channel pairs as a function of the similarity threshold. For strict thresholds (e.g., $\tau \ge 0.90$), no coordinated pairs are detected, indicating that near-duplicate, near-simultaneous posting across mainstream news channels is extremely rare.

\subsection{Daily Coordination ($\texttt{D}$ buckets)}

\begin{figure}[H]
    \centering
    \includegraphics[width=0.7\linewidth]{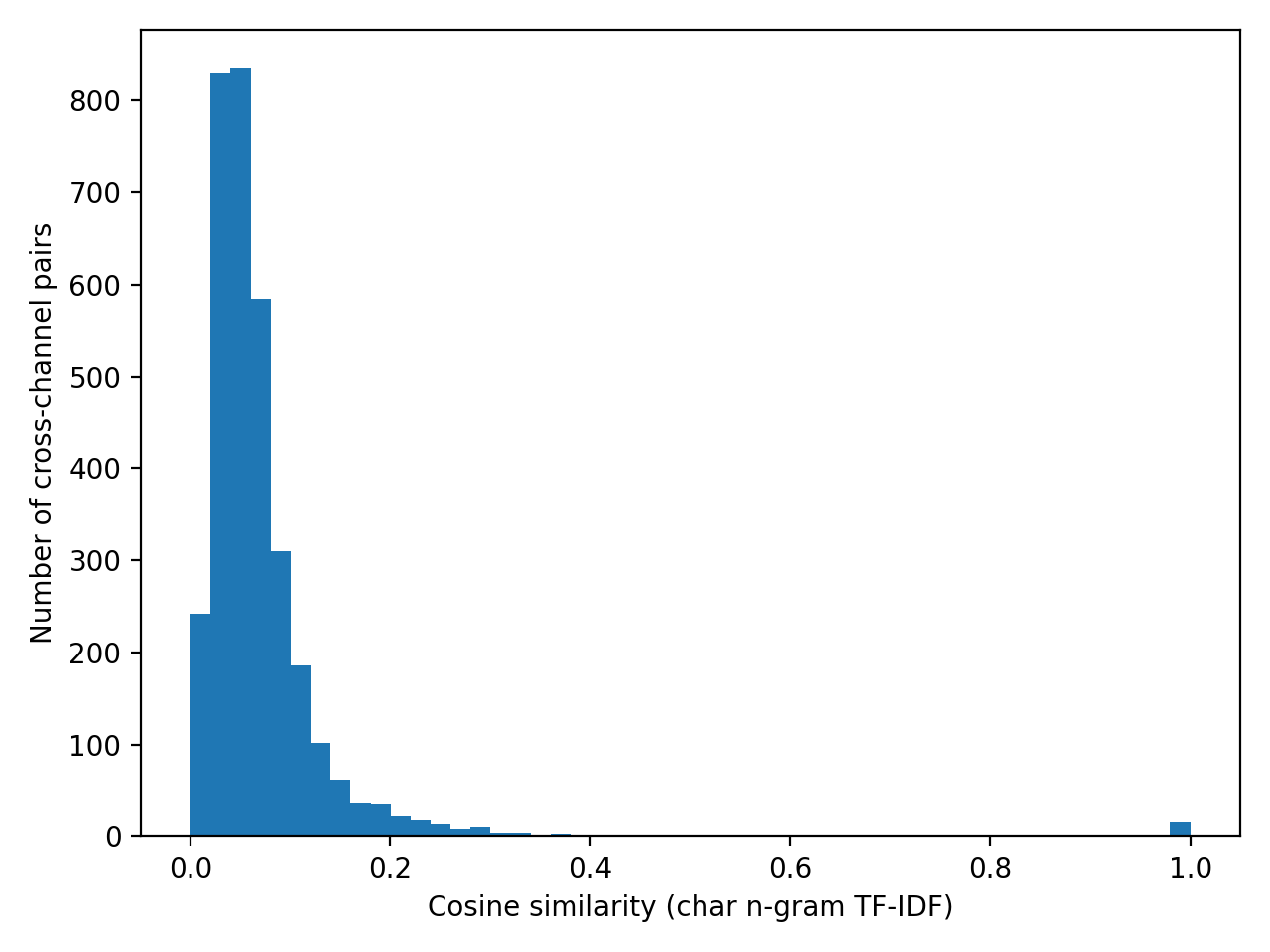}
    \caption{Distribution of cross-channel cosine similarities within daily time buckets ($\texttt{D}$). Daily aggregation reveals limited same-day textual alignment across channels.}
    \label{fig:sim_hist_D}
\end{figure}

To examine looser temporal alignment, we repeat the analysis using daily time buckets. Fig.~\ref{fig:sim_hist_D} shows the similarity distribution under daily aggregation.

While the distribution remains skewed toward low similarity values, a slightly larger tail of moderately high similarity scores emerges compared to the hourly setting. This suggests that some channels publish similar content on the same day, even if not within the same hour.

\begin{figure}[H]
    \centering
    \includegraphics[width=0.7\linewidth]{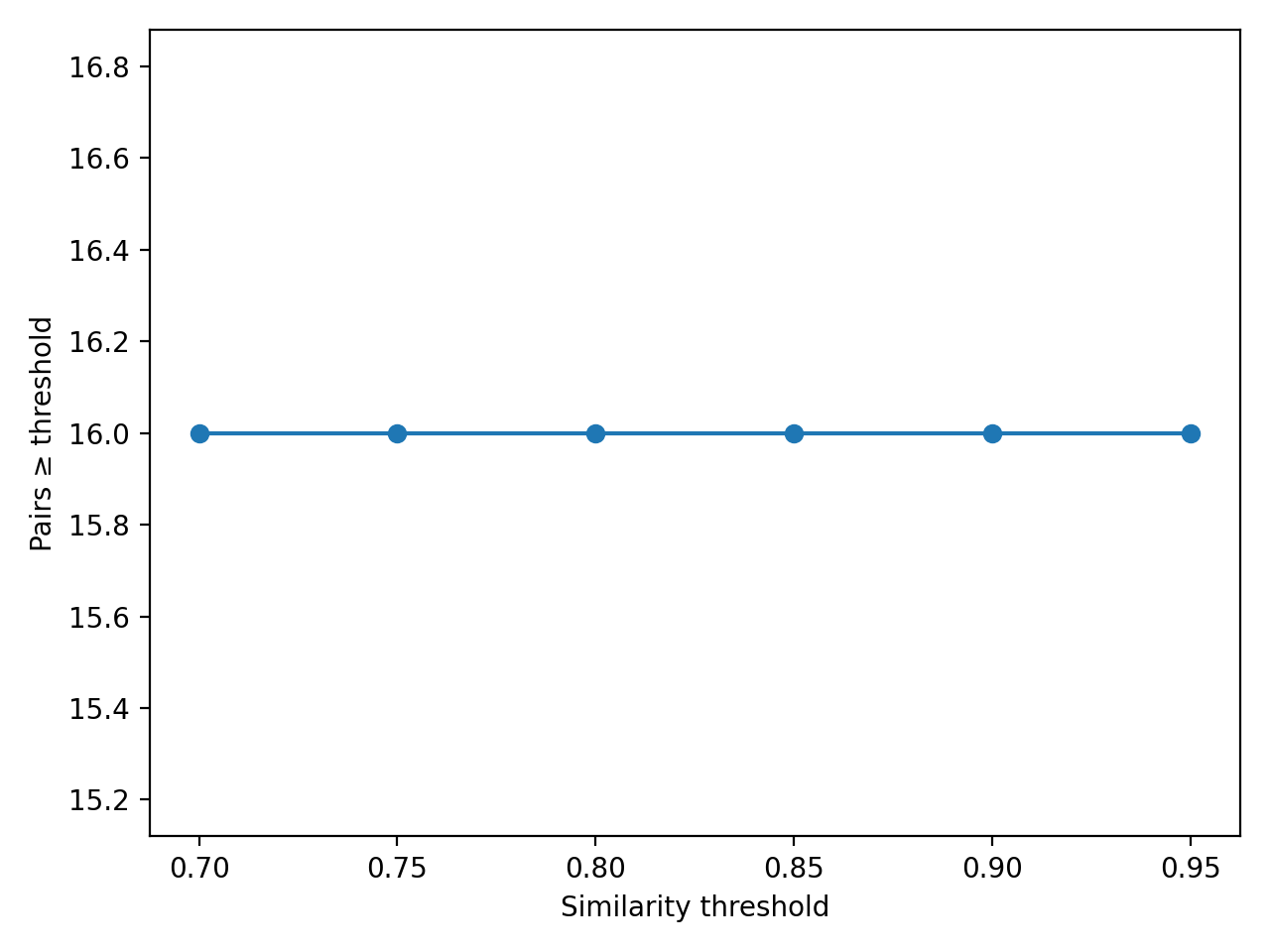}
    \caption{Number of detected cross-channel pairs as a function of the similarity threshold under daily bucketing. A small number of same-day near-duplicate pairs emerge for moderate thresholds.}
    \label{fig:pairs_vs_threshold_D}
\end{figure}

Fig.~\ref{fig:pairs_vs_threshold_D} shows that a small number of cross-channel pairs are detected for moderate thresholds (e.g., $\tau \ge 0.85$). These cases correspond to same-day reporting of major events rather than strict temporal synchronization.

\subsection{Negative Control: Timestamp Randomization}
\label{subsec:negative_control}
\justifying

To validate that the coordination detection framework does not produce spurious alignment, we conducted a negative control experiment in which message timestamps were randomly permuted within each Telegram channel. This preserves per-channel message volume and textual content while destroying any real temporal synchronization across channels. We then re-ran the full coordination detection pipeline using the same similarity thresholds and temporal bucketing as in the main analysis.

\begin{figure}[H]
    \centering
    \includegraphics[width=0.7\linewidth]{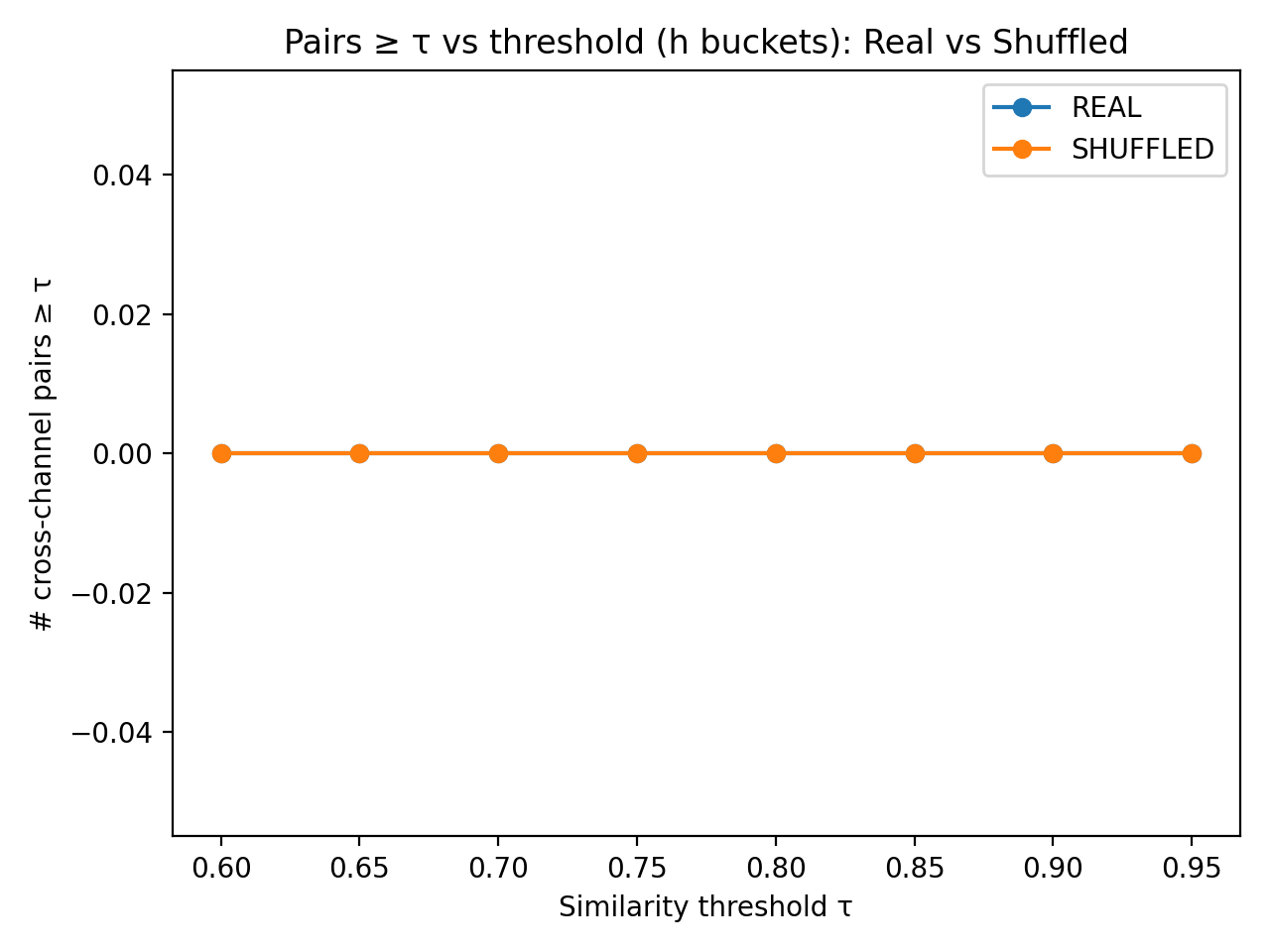}
    \caption{Negative control using timestamp randomization. The number of detected cross-channel pairs remains zero for both the real and shuffled data across all similarity thresholds, demonstrating that the method does not generate spurious coordination.}
    \label{fig:neg_control_pairs_h}
\end{figure}

Across all tested thresholds, the negative control yields zero detected coordinated pairs, matching the results obtained on the original data. Fig.~\ref{fig:neg_control_pairs_h} shows that the number of detected cross-channel pairs remains at zero for both the real and shuffled data across the full range of similarity thresholds. This result confirms that the proposed methodology is conservative and does not introduce artificial coordination signals; the absence of detected coordination in the original data therefore reflects genuine structural properties of mainstream news dissemination on Telegram rather than methodological artifacts.

\subsection{Interpretation}
\justifying

Across both temporal resolutions, coordination among mainstream global news channels remains limited. Hourly analysis reveals virtually no near-duplicate synchronization, while daily analysis captures only a small number of same-day similarities. Qualitative inspection suggests that these cases correspond to routine journalistic coverage of major events rather than sustained or strategic coordinated dissemination. Overall, these findings reinforce the conclusion that coordinated information dissemination on Telegram, when measured conservatively through temporal alignment and near-duplicate textual similarity, is rare in mainstream global news ecosystems.

\subsection{External Context: Major Events Around Peak Volumes}

Temporal analysis reveals a prominent surge in Venezuela-related message volume beginning on 3 January 2026, corresponding with a significant geopolitical event: on that date the United States launched military strikes in Venezuela and announced the capture of President Nicolás Maduro and his wife, Cilia Flores, subsequently flying them to the United States on federal charges. This event was widely covered by global media and prompted international debate regarding sovereignty and international law. Continued elevated volume in early January reflects follow-on political developments, including the swearing-in of an interim president, diplomatic reactions at the United Nations, and public protests in allied nations such as Cuba. These external events provide real-world anchors for understanding peaks in the Telegram news volume data, even though our coordination analysis did not reveal near-duplicate synchronization aligned with them.

\begin{figure}
    \centering
    \includegraphics[width=0.9\linewidth]{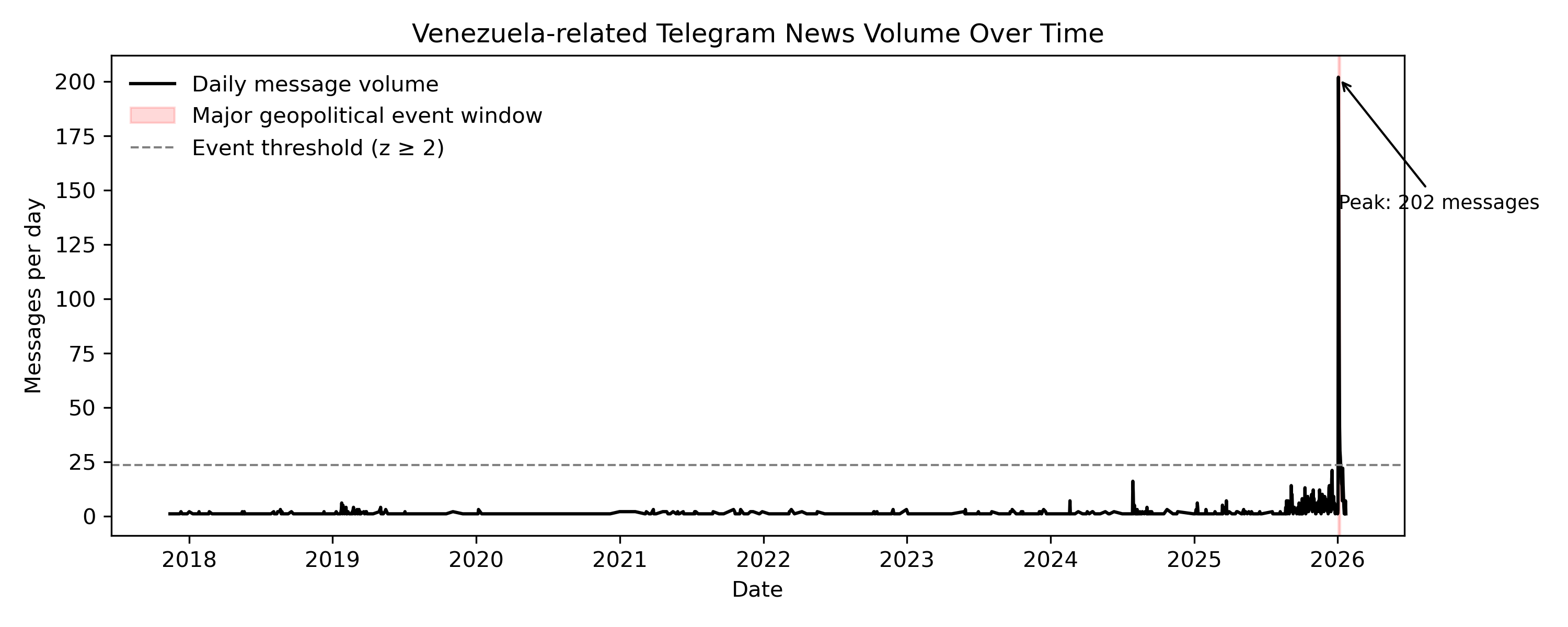}
\caption{Daily volume of Venezuela-related messages published by mainstream Telegram news channels over time. A pronounced spike in early January 2026 corresponds to a major geopolitical escalation involving U.S. military action in Venezuela. Despite the extreme increase in reporting activity, coordination analysis reveals no near-duplicate textual synchronization among channels, indicating event-driven attention rather than coordinated content dissemination.}
\label{fig:fig_event_volume}
\end{figure}
\paragraph{Event-driven volume without coordination.}
Fig.~\ref{fig:fig_event_volume} illustrates the daily volume of Venezuela-related Telegram messages over the full observation period, highlighting a pronounced and abrupt spike in early January 2026. This surge corresponds to a major geopolitical escalation involving direct U.S. military intervention in Venezuela and the capture of President Nicolás Maduro. The magnitude of this peak substantially exceeds baseline activity levels, with message volume increasing by more than an order of magnitude relative to typical daily counts.

Importantly, this volume spike reflects a rapid and widespread increase in editorial attention across international news outlets rather than coordinated content reuse. Despite the temporal concentration of reporting, channels largely published distinct articles, headlines, and narrative framings. This observation is consistent with our coordination analysis results, which detect no near-duplicate textual synchronization during this period, even under looser daily temporal aggregation.

The contrast between extreme volume and the absence of detected coordination underscores a key distinction between \emph{attention synchronization} and \emph{content coordination}. Major geopolitical events naturally induce simultaneous coverage across outlets; however, such simultaneity does not necessarily imply coordinated messaging or shared editorial control. In this case, mainstream news organizations responded independently to the same external trigger, producing heterogeneous textual content despite overlapping publication times.

This finding reinforces the conservative nature of our coordination detection methodology. Even under conditions of intense global news pressure and heightened reporting frequency, the method does not falsely infer coordination in the absence of near-duplicate textual similarity. Consequently, the lack of detected coordination during this event-driven surge should be interpreted as a substantive result rather than a limitation of the analytical framework.

\subsection{Per-Channel Activity During the January 2026 Peak}

To better understand the drivers of the January 2026 volume spike, we examine the per-channel contribution to message volume during this period. Fig.~\ref{fig:fig_jan2026_channel_volume} shows daily Venezuela-related message counts for each Telegram channel between 1 and 10 January 2026.
The surge is primarily driven by a small subset of high-volume international outlets, with RT News contributing the majority of messages, followed by BBC World and France 24. Other channels exhibit sporadic or minimal activity, reflecting editorial prioritization rather than synchronized behavior.

Temporal inspection further reveals a staggered publication pattern across channels rather than near-simultaneous posting. While some outlets report within the same day, message timestamps are distributed over several hours, consistent with independent news production workflows.

These findings reinforce the coordination analysis results: despite intense and concentrated coverage of a major geopolitical event, the observed behavior reflects event-driven attention rather than coordinated or scripted information dissemination.

\subsection{Per-channel dynamics during peak activity (January 2026)}
\begin{figure}
    \centering
    \includegraphics[width=0.8\linewidth]{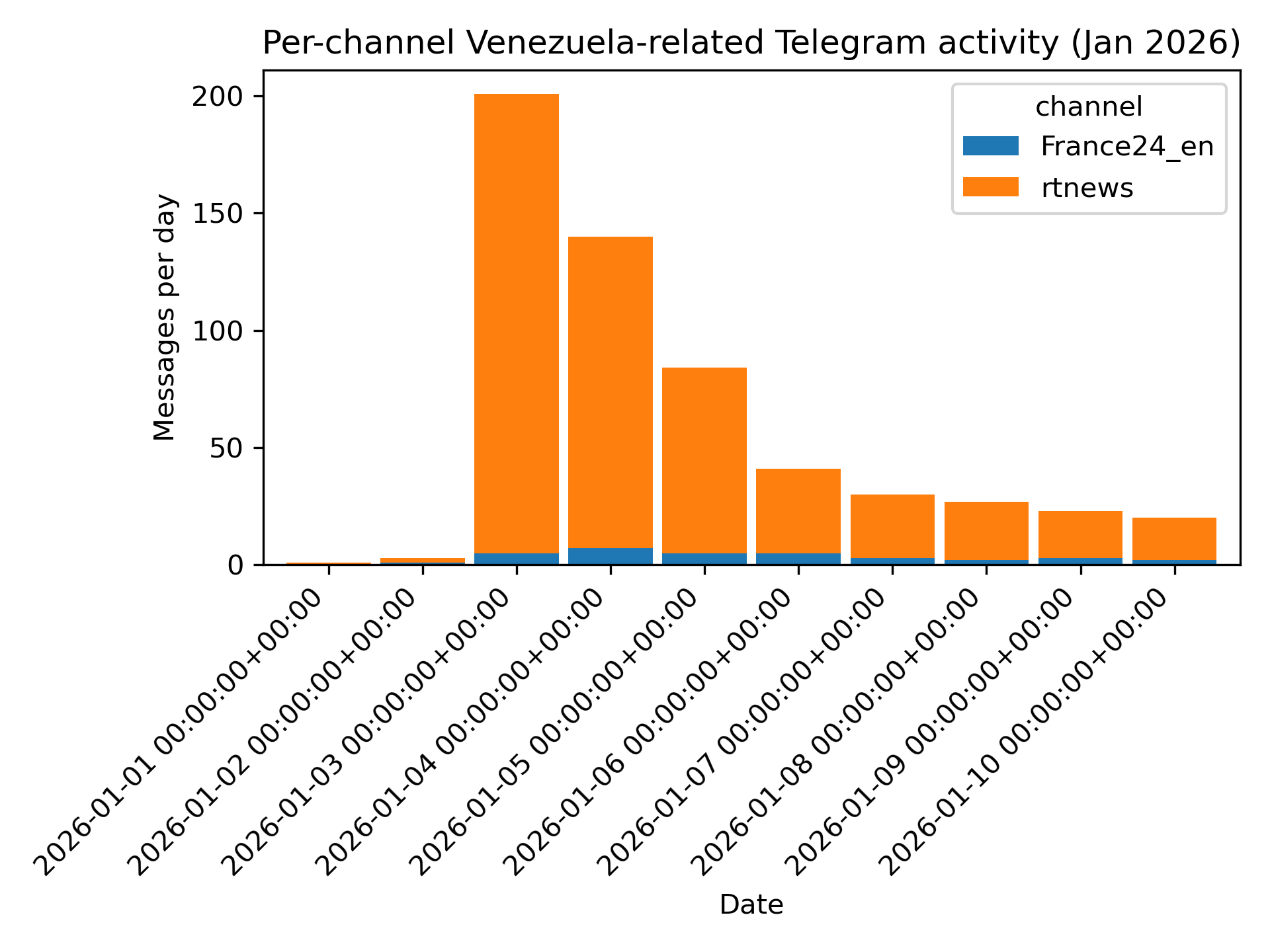}
    \caption{ Per-channel distribution of Venezuela-related Telegram messages during
January 2026. The sharp increase in activity is driven primarily by RT News,
with limited contribution from other international outlets. Despite the volume
surge, no cross-channel coordination is detected. 
}
    \label{fig:fig_jan2026_channel_volume}
\end{figure}

Fig.~\ref{fig:fig_jan2026_channel_volume} presents a per-channel breakdown of
Venezuela-related Telegram activity during the peak period of January 2026.
The surge begins abruptly on 3 January, reaches a maximum on 4 January, and
gradually decays over the following days. This temporal pattern is consistent
with a breaking-news cycle, characterized by an initial shock phase followed by
progressively reduced editorial intensity.

The increase in volume is overwhelmingly driven by a single outlet (RT News),
which accounts for the majority of messages throughout the peak period. Other
international news channels contribute only sporadically, indicating that the
observed volume spike reflects asymmetric editorial focus rather than collective
synchronization across outlets.

Importantly, despite the sharp increase in message volume, our coordination
analysis does not detect near-duplicate content or cross-channel temporal
alignment. This suggests that the surge corresponds to independent reporting on
the same underlying developments, rather than coordinated information
dissemination. The figure thus illustrates a key distinction between volume-based
attention peaks and coordinated messaging behavior.

\section{Discussion: Disentangling Attention and Coordination in Telegram News Coverage}
\justifying

This section synthesizes the coordination results with complementary analyses of temporal leadership and narrative diversity. The goal is to distinguish synchronized attention (simultaneous reporting intensity) from coordination in content (near-duplicate reuse).

This study examined whether heightened geopolitical attention on Telegram is accompanied by coordinated information dissemination among mainstream international news outlets. By jointly analyzing temporal activity patterns, near-duplicate textual similarity, lead--lag dynamics, and narrative clustering, we provide converging evidence that synchronization in attention does not imply coordination in content.

First, extreme volume spikes—most notably during early January 2026—reflect rapid, event-driven editorial responses rather than synchronized messaging behavior. Despite near-simultaneous reporting across channels, we observe no near-duplicate textual reuse under either hourly or daily temporal aggregation. Negative control experiments based on timestamp randomization further confirm that the coordination detection framework does not generate spurious signals, even under conditions of intense reporting pressure. These findings indicate that high-volume, concurrent coverage alone is insufficient evidence of coordinated dissemination.

Second, lead--lag analysis reveals pronounced temporal asymmetry across channels. A small subset of high-volume outlets consistently acts as early reporters within short-lived attention windows, while others follow sporadically. Crucially, this temporal hierarchy does not correspond to near-duplicate or coordinated content reuse. Instead, it reflects heterogeneous editorial responsiveness and agenda-setting behavior, reinforcing the distinction between temporal attention alignment and coordinated dissemination strategies.

Third, narrative clustering during the January 3–6, 2026 peak reveals moderate thematic diversity without clearly separable narrative blocs. As shown in Fig.~\ref{fig:narrative_scatter}, posts form a dense and overlapping semantic space in the reduced-dimensional projection, rather than distinct or polarized clusters. Cluster size distributions further support this interpretation, with two dominant clusters accounting for most messages and several smaller clusters capturing peripheral subtopics. The overall narrative entropy (1.87) indicates limited but non-trivial diversity, consistent with mainstream journalistic coverage rather than fragmented or adversarial framing.

Channel-level entropy analysis reveals strong asymmetries in narrative breadth. France24 exhibits low narrative entropy (0.83), with most content concentrated in a single cluster, reflecting a narrow editorial focus during the peak period. In contrast, RT News spans all identified clusters and displays entropy comparable to the overall dataset (1.89), indicating broad thematic coverage. Importantly, this diversity arises from volume dominance and editorial breadth rather than coordinated alignment with other outlets.

Taken together, these results highlight a critical distinction between \emph{attention synchronization} and \emph{content coordination}. Major geopolitical events naturally induce simultaneous coverage across news organizations; however, such simultaneity does not necessitate shared narratives or coordinated dissemination. In this mainstream news ecosystem, Telegram functions primarily as a broadcast medium that supports rapid, independent reporting rather than synchronized narrative control.

From a methodological perspective, the comparison with Reddit underscores that coordination detection is inherently platform-dependent. While Telegram's channel-based architecture permits robust falsification tests, sensitivity analyses, and narrative-level validation, structural sparsity on Reddit renders analogous coordination experiments infeasible. This asymmetry emphasizes the importance of aligning coordination metrics with platform affordances and data-generating processes.

Overall, our findings establish a conservative empirical baseline for mainstream news behavior on Telegram. The absence of detected coordination under extreme attention conditions suggests that coordinated information dissemination—when it occurs—is more likely to emerge in partisan, state-aligned, or activist ecosystems rather than in professional international journalism.
\begin{comment}
    \begin{figure}
    \centering
    \includegraphics[width=0.5\linewidth]{clusters_scatter.png}
    \caption{  Narrative clustering (k=5) of Venezuela-related Telegram posts during January 3–6, 2026, projected onto the first two SVD components. Colors denote cluster assignments. The absence of clearly separable clusters and the strong overlap between narratives indicate moderate thematic variation without polarized or coordinated narrative blocs.
}
    \label{fig:clusters}
\end{figure}

\end{comment}

\subsection{Narrative Diversity During the January 2026 Peak}
\begin{figure}[H]
    \centering
    \includegraphics[width=0.70\linewidth]{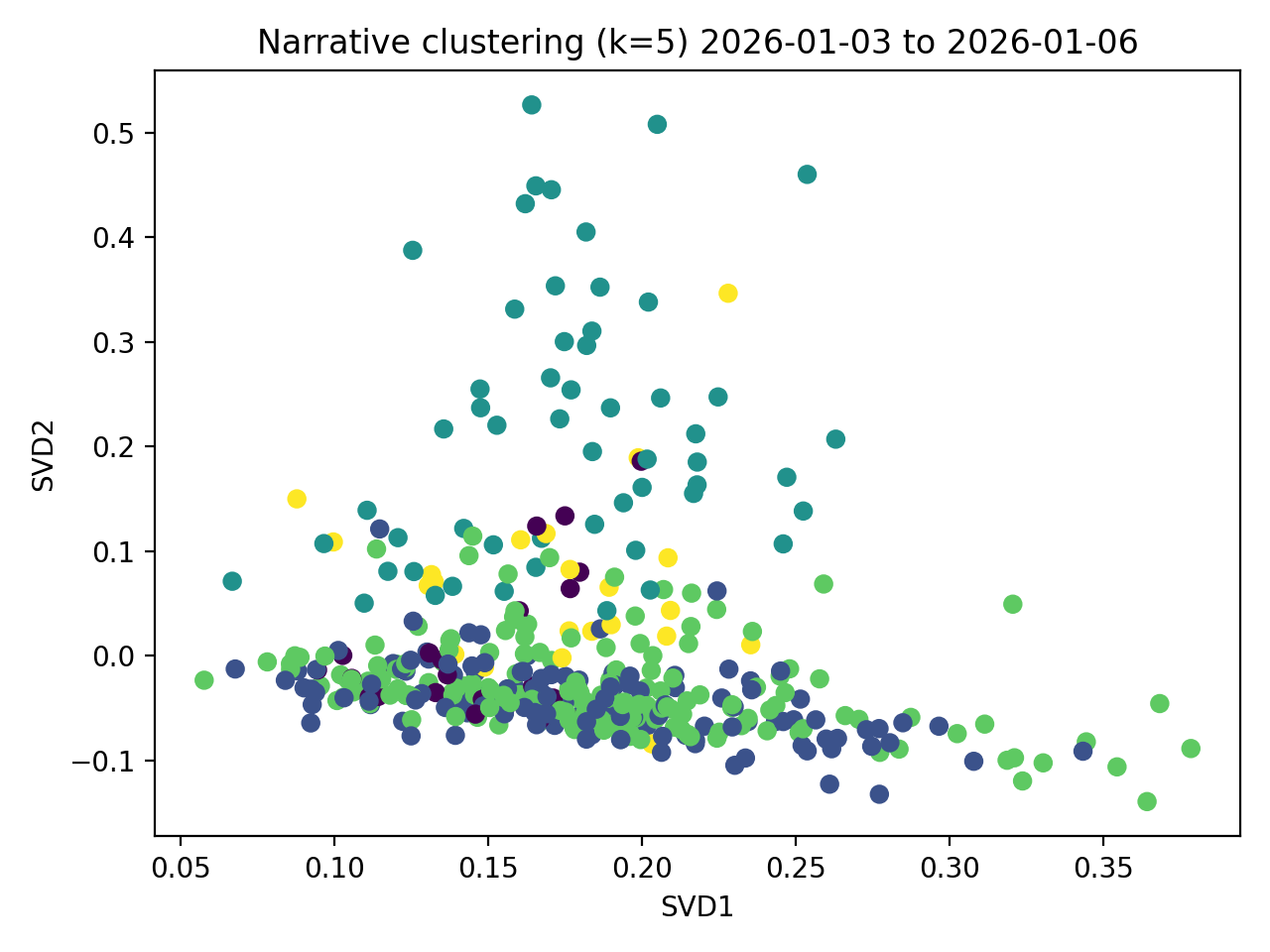}
    \caption{Narrative clustering of Venezuela-related Telegram posts during 3--6 January 2026 ($k=5$), projected onto the first two SVD components. Colors denote cluster assignments. Strong overlap between clusters suggests heterogeneous framing and the absence of separable narrative blocs during the peak attention period.}
    \label{fig:narrative_scatter}
\end{figure}

\justifying
To assess whether heightened attention corresponds to convergent framing, we cluster messages posted during the peak window (3--6 January 2026) using character $n$-gram TF--IDF representations and $k$-means ($k=5$). Figure~\ref{fig:narrative_scatter} visualizes the resulting narrative clusters under a 2D SVD projection. The clusters overlap substantially and do not form clearly separable blocs, indicating moderate thematic variation rather than polarized or coordinated narrative alignment across outlets.

\subsection{Attention--Coordination Ratio}
\begin{figure}[H]
    \centering
    \includegraphics[width=0.70\linewidth]{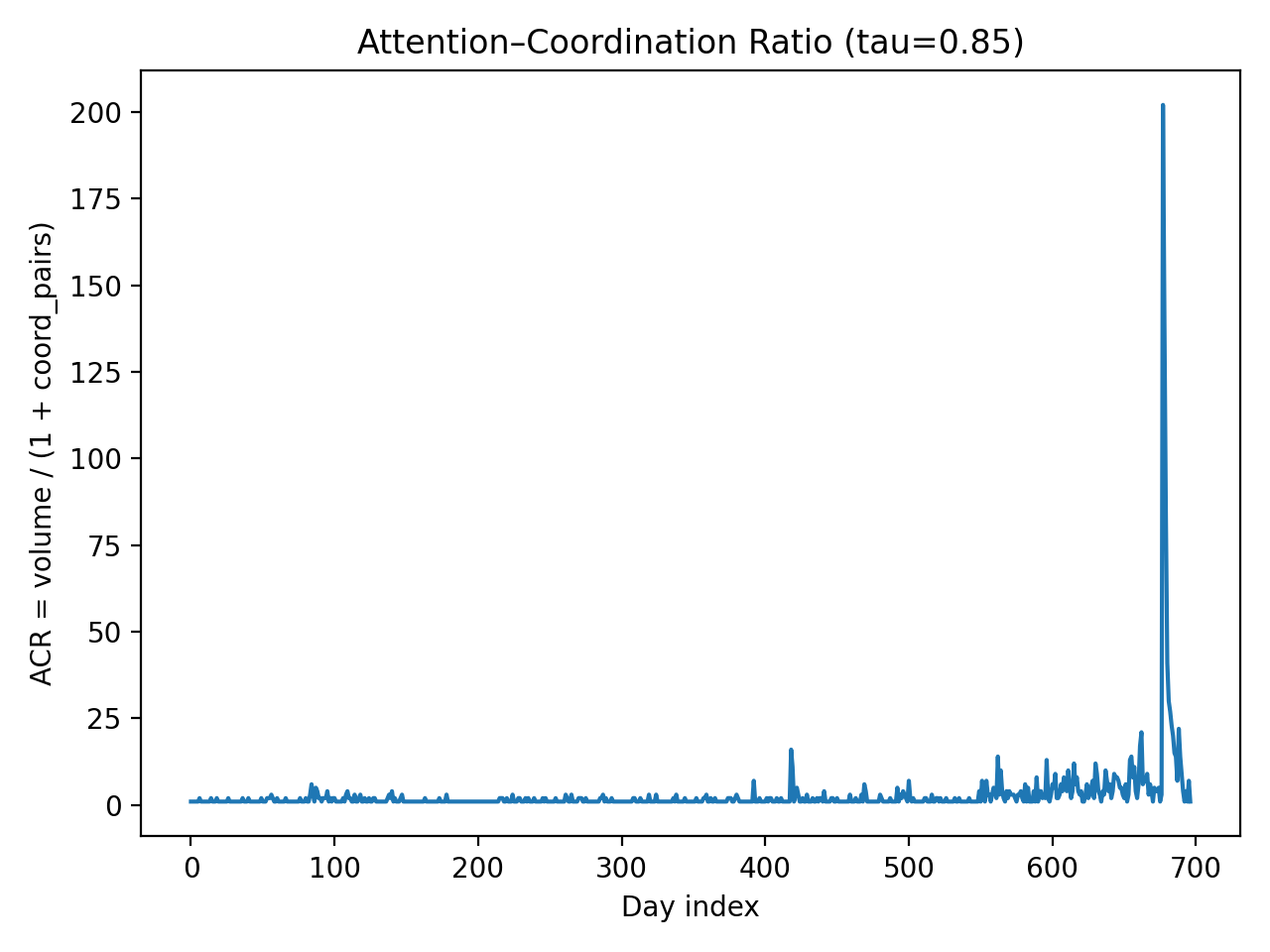}
    \caption{Attention--Coordination Ratio (ACR) time series for $\tau=0.85$. Large values indicate high volume in the absence of detected near-duplicate coordination, highlighting attention synchronization without content coordination.}
    \label{fig:acr}
\end{figure}

\justifying
To explicitly disentangle attention peaks from coordination, we define an \emph{Attention--Coordination Ratio (ACR)} per day as:
\[
\mathrm{ACR}(d) = \frac{\mathrm{Volume}(d)}{1+\mathrm{CoordPairs}(d)} ,
\]
where $\mathrm{Volume}(d)$ is the number of Venezuela-related posts and $\mathrm{CoordPairs}(d)$ is the number of cross-channel near-duplicate pairs detected on day $d$ at threshold $\tau=0.85$. The $+1$ term avoids division by zero. Figure~\ref{fig:acr} shows that ACR becomes extremely large during the January 2026 spike, reflecting a sharp increase in attention while coordinated near-duplicate reuse remains absent.

\paragraph{Coordination network projection.}
Projecting detected coordination events into a channel--channel graph yields a very sparse network at $\tau=0.85$, reflecting the limited number of cross-channel near-duplicate pairs under the conservative coordination definition.

\begin{comment}
\end{comment}

\section{Ethical Considerations}
\justifying

All analyzed data originate from publicly accessible Telegram channels. No private messages, user identifiers, or personal data were collected or processed. The analysis adheres to established ethical guidelines for computational social science research and minimizes potential risks to individuals or groups.

\section{Acknowledgments}
\justifying

This research was supported by MEDIATE project (101074075 GA) funded by the European Union.
\section*{Appendix A: Auxiliary Reddit Coordination Experiments}
\begin{figure}[H]
    \centering
    \includegraphics[width=0.55\linewidth]{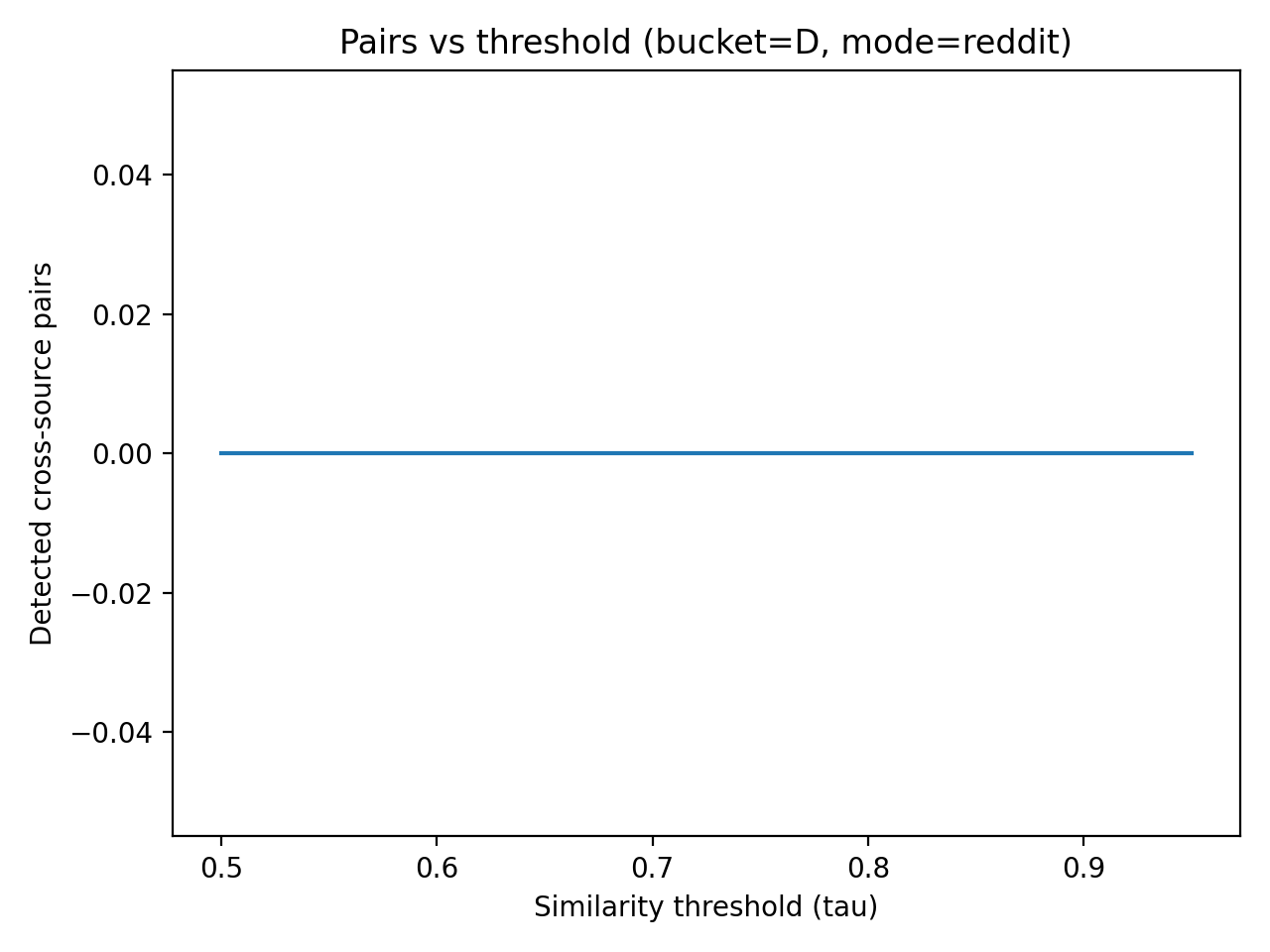}
    \caption{Sensitivity analysis of cross-subreddit coordination on Reddit. The number of detected cross-source pairs remains zero for all similarity thresholds $\tau$, reflecting the absence of comparable temporal buckets rather than threshold-dependent effects.}
    \label{fig:reddit_sensitivity}
\end{figure}
\paragraph{Cross-source coordination network.}
Projecting detected coordination events onto a subreddit--subreddit graph yields an empty network (0 nodes, 0 edges) for $\tau=0.85$, as no cross-subreddit near-duplicate pairs exist within shared temporal windows.

\begin{figure}[H]
    \centering
    \includegraphics[width=0.6\linewidth]{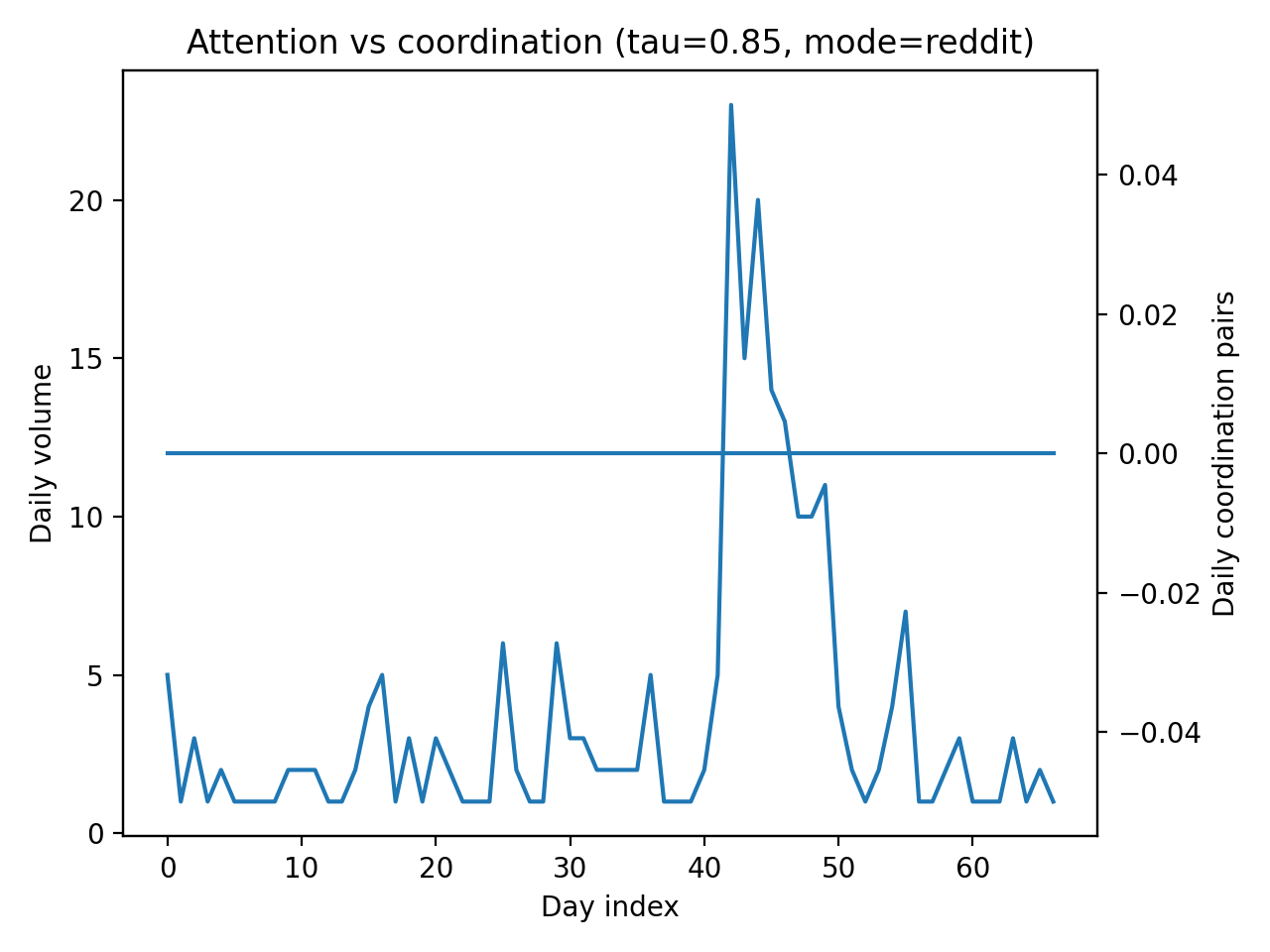}
    \caption{Daily Reddit submission volume versus detected coordination pairs ($\tau=0.85$). While attention varies over time, the coordination signal remains identically zero due to the absence of cross-subreddit temporal overlap.}
    \label{fig:reddit_attention_coord}
\end{figure}

\bibliography{main}

\end{document}